\documentclass[fleqn,usenatbib]{mnras}

\usepackage{newtxtext,newtxmath}

\usepackage[T1]{fontenc}
\usepackage{ae,aecompl}

\newcommand{\OmegaB}{\Omega_{\rm B}}
\newcommand{\OmegaM}{\Omega_{\rm M}}
\newcommand{\OmegaL}{\Omega_{\Lambda}}

\newcommand{\Rvir}{R_{\rm vir}}
\newcommand{\vvir}{v_{\rm vir}}
\newcommand{\rhovir}{\rho_{\rm vir}}
\newcommand{\Tvir}{T_{\rm vir}}
\newcommand{\rhoM}{\bar{\rho }_{\rm M}}
\newcommand{\MSUN}{\mathrm{M_{\sun}}}

\newcommand{\fB}{f_{\rm B}}

\newcommand{\rhoB}{\bar{\rho}_{\rm B}}

\newcommand{\kB}{k_{\mathrm{B}}}
\newcommand{\cs}{c_{\mathrm{s}}}

\newcommand{\Mgas}{M_{\mathrm{gas}}}
\newcommand{\rhogas}{\rho_{\mathrm{gas}}}

\newcommand{\deltavir}{\Delta_{\rm vir}}
\newcommand{\Z}{\left( \dfrac{1+z}{10}\right)}
\newcommand{\nH}{n_{\rm H}}
\newcommand{\tff}{t_{\mathrm{ff}}}
\newcommand{\tcool}{t_{\mathrm{cool}}}
\newcommand{\fmol}{f_{\rm H_{2}}}
\newcommand{\J}{\bar{J}_{\rm LW, 21}}

\newcommand{\HI}{H\,{\sc i}~}
\newcommand{\HII}{H\,{\sc ii}~}

\usepackage{xcolor}

\usepackage{graphicx}	
\usepackage{amsmath}	

\title[The critical halo mass for star formation]{Starbursts in low-mass haloes at Cosmic Dawn. I. The critical halo mass for star formation}

\author[O. Nebrin et al.]{
\parbox[t]{\textwidth}{
Olof Nebrin$^1$\thanks{Email: olof.nebrin@astro.su.se}, Sambit K. Giri$^2$, Garrelt Mellema$^1$}
\vspace*{6pt} \\
$^1$ Department of Astronomy \& Oskar Klein Centre, AlbaNova, Stockholm University, SE-106 91 Stockholm, Sweden\\
$^{2}$ Nordita, KTH Royal Institute of Technology and Stockholm University, Hannes Alf\'vens v\"ag 12, SE-106 91 Stockholm, Sweden
}

\date{Accepted XXX. Received YYY; in original form ZZZ. NORDITA-2023-008}

\pubyear{2023}

\begin{document}
\label{firstpage}
\pagerange{\pageref{firstpage}--\pageref{lastpage}}
\maketitle

\begin{abstract}
The first stars, galaxies, star clusters, and direct-collapse black holes are expected to have formed in low-mass ($\sim$$10^{5}-10^{9} ~ \MSUN$) haloes at Cosmic Dawn ($z \sim 10 - 30$) under conditions of efficient gas cooling, leading to gas collapse towards the centre of the halo. The halo mass cooling threshold has been analyzed by several authors using both analytical models and numerical simulations, with differing results. Since the halo number density is a sensitive function of the halo mass, an accurate model of the cooling threshold is needed for (semi-)analytical models of star formation at Cosmic Dawn. In this paper the cooling threshold mass is calculated (semi-)analytically, considering the effects of H$_2$-cooling and formation (in the gas phase and on dust grains), cooling by atomic metals, Lyman-$\alpha$ cooling, photodissociation of H$_2$ by Lyman-Werner photons (including self-shielding by H$_2$), photodetachment of H$^-$ by infrared photons, photoevaporation by ionization fronts, and the effect of baryon streaming velocities. We compare the calculations to several high-resolution cosmological simulations, showing excellent agreement. We find that in regions of typical baryon streaming velocities, star formation is possible in haloes of mass $\gtrsim 1-2 \times 10^6 ~ \MSUN$ for $z \gtrsim 20$. By $z \sim 8$, the expected Lyman-Werner background suppresses star formation in all minihaloes below the atomic-cooling threshold ($\Tvir = 10^4 ~ \textrm{K}$). The halo mass cooling threshold increases by another factor of $\sim$$4$ following reionization, although this effect is slightly delayed ($z \sim 4-5$) because of effective self-shielding.

\end{abstract}

\begin{keywords}
galaxies: formation - intergalactic medium - cosmology: dark matter - dark ages, reionization, first stars
\end{keywords}



\section{Introduction}

In the standard $\Lambda$CDM paradigm, structure form in a bottom-up manner, with low-mass haloes virializing first, and subsequently merging to form larger haloes. Star formation can only take place in those haloes wherein gas can cool efficiently, lose its pressure support, and collapse to high densities \citep[e.g.][]{WhiteRees1978, Blumenthal1984, Tegmark1997, TegmarkRees1998, Nishi1999, Tegmark2006, Bromm2013, Liu2019}. The very first stars are therefore expected to form in the smallest haloes where efficient gas cooling is possible. These haloes have virial temperatures $\Tvir < 10^4 ~ \rm K$ and are known as minihaloes, and the dominant cooling agent within them is molecular hydrogen, H$_{2}$. Subsequent radiative feedback from the first stars -- most notably Lyman-Werner (LW) radiation ($11.2-13.6 ~ \rm eV$ photons)  -- can suppress H$_2$-cooling in minihaloes, potentially raising the minimum mass for efficient cooling to haloes with $\Tvir \simeq 10^4 ~ \rm K$, above which atomic line cooling becomes effective.

The halo mass function at high redshifts is a sensitive function of the halo mass $M$, and because of this, uncertainties in the assumed mass threshold for efficient cooling (henceforth the cooling threshold) can propagate into large uncertainties in the predicted abundance and characteristics of the first stars and galaxies in semi-analytical models \citep[e.g.][]{Trenti2009, Gao2010, Fialkov2012, Crosby2013, Griffen2018, Mebane2018, Mebane2020, Munos2022, Hartwig2022}. The cooling threshold has been investigated by several authors over the years using (semi-)analytical calculations \citep[e.g.][]{Blumenthal1984, Tegmark1997, Abel1998_minimummass, Nishi1999, Trenti2009, Visbal2014_H2, Liu2019, Lupi2021},\footnote{\citet{Blumenthal1984} appear to have provided the first analytical expression for the cooling threshold in the context of cold dark matter (CDM) cosmology, utilizing early H$_2$-cooling rates from \citet{Yoneyama1972}. Several other authors studied the cooling of collapsing gas clouds in an expanding Universe, but not in the context of CDM cosmology \citep[e.g.][]{Hirasawa1969, Matsuda1969, Yoneyama1972, Hutchins1976, Lahav1986}.} idealized non-cosmological simulations \citep[e.g.][]{Haiman1996, Fuller2000, Nakatani2020}, and fully cosmological hydrodynamical simulations \citep[e.g.][]{Machacek2001, Yoshida2003, Latif2019_cooling, Schauer2019, Skinner2020, Kulkarni2021, Park2021, Schauer2021}. 
Despite a large number of studies on the cooling threshold, substantial differences in its value can be found. Some of this can be traced to the adoption of outdated cooling rates \citep[e.g.][]{Machacek2001}, too low gas mass resolution \citep[e.g.][]{Hummel2012}, or the neglect of self-shielding when studying the effect of LW feedback \citep[e.g.][]{Machacek2001, Trenti2009}. A summary of the results of studies on the cooling threshold and their potential drawbacks is presented in Table \ref{TableFitMcrit}. 

Recently, \citet{Schauer2019, Schauer2021} and \citet{Kulkarni2021} independently studied the halo mass cooling threshold using high-resolution cosmological hydrodynamical simulations, both with and without LW feedback and baryon streaming velocities. Both teams used up-to-date chemistry and cooling rates, and also took self-shielding of H$_2$ into account. Both teams found that the H$_2$-cooling threshold is not sharp --- there is a minimum halo mass $M_{\rm min}$ below which no haloes host cool and dense gas. The fraction of haloes hosting cool gas steadily increases for $M > M_{\rm min}$, until it saturates near unity. To study the \textit{typical} halo mass at gas collapse, \citet{Schauer2019,Schauer2021} focus on $M_{\rm ave}$, the median halo mass at collapse, whereas \citet{Kulkarni2021} focus on $M_{\rm 50\%}$, the halo mass above which $>50\%$ of haloes host cool and dense gas. In practice, all else being equal we expect that $M_{\rm 50\%} \simeq M_{\rm ave}$. Despite this, the results of these two studies are in tension with one another. In particular, \citet{Schauer2019, Schauer2021} found values of $M_{\rm min}$  that were a factor $\sim 2$ larger than those found by \citet{Kulkarni2021}. The discrepancy is larger for the "typical" halo mass at collapse --- $M_{\rm ave}$ in \citet{Schauer2019, Schauer2021} is a factor $\sim 3-4$ larger than $M_{\rm 50\%}$ in \citet{Kulkarni2021} (see Figure \ref{H2coolingfigure}). The reason(s) for this discrepancy is not entirely clear and under investigation by both teams, but \citet{Schauer2021} suggests that this could be due to different halo mass definitions and/or collapse criteria. 

In cosmological simulations, the exact critical halo mass for star formation is of limited interest --- if stars form in two haloes, there will be two luminous objects regardless of how the masses of the haloes are estimated. This luxury of independence is unfortunately not available in semi-analytical models, which has led to researchers having to choose between the results from \citet{Schauer2019,Schauer2021} and \citet{Kulkarni2021} in semi-analytical models \citep[e.g.][]{Wu2021, Hartwig2022, Munos2022, Uysal2023, Ventura2023}.

\begin{table*}
\caption{The H$_2$-cooling threshold with and without Lyman-Werner (LW) feedback: Results of prior studies and their potential limitations. Here $z_{10} \equiv (1+z)/10$, $z_{21} \equiv (1+z)/21$, $\J \equiv \bar{J}_{\rm LW}/10^{-21} ~ \mathrm{erg} ~ \mathrm{s}^{-1} ~ \mathrm{ cm}^{-2} ~ \mathrm{ Hz}^{-1} ~\mathrm{ sr}^{-1}$ is the normalized frequency-averaged intensity in the LW band (11.2-13.6 eV), and $J_{21} \equiv J_{\nu}(h\nu = 13.6 ~ \textrm{eV})/10^{-21} ~ \mathrm{erg} ~ \mathrm{s}^{-1} ~ \mathrm{ cm}^{-2} ~ \mathrm{ Hz}^{-1} ~\mathrm{ sr}^{-1}$ is the normalized intensity at $h\nu = 13.6 ~ \rm eV$, at the edge of the LW band \citep[e.g.][]{Safranek2012_LW}. The two intensities are related by $\J = \beta J_{21}$, where $\beta$ depends on the assumed spectrum \citep[e.g.][]{Agarwal2015}. For a $10^5 ~ \rm K$ blackbody spectrum \citep[assumed by][]{Schauer2021} $\beta \simeq 0.905$, which increases to $\beta \simeq 1.56$ and $\beta \simeq 3.68$ for $2 \times 10^4~ \rm K$ \citep[assumed by][]{Latif2019_cooling} and $10^4~ \rm K$ blackbody spectra, respectively.$^{\diamond}$}
\begin{tabular}{l l l l l}
\hline
\hline
Reference & $\J$ or $J_{21}$ & Method & Cooling threshold ($10^6 ~ \MSUN$) & Potential limitation(s) \\

\hline
\hline

\\

 \citet{Tegmark1997} & $\J = 0$ & Semi-analytical & $M_{\rm min} \sim 1-10$ (no fit provided) & Outdated cooling rate$^{\dagger}$\\ 
 
 \\
 
 \citet{Abel1998_minimummass} & $\J = 0$ & Semi-analytical & $M_{\rm min} \sim 0.1-10$ (no fit provided) & Outdated cooling rate$^{\dagger}$\\ 
 
 \\
 
 \citet{Fuller2000} & $\J = 0$ & Simulations & $M_{\rm min} = 2.0 \hspace{1 pt} z_{10}^{-2.0}$ & Einstein-de Sitter cosmology \\
 
 \\
 
 \citet{Machacek2001} & $\J = 0-0.08$ & Simulations & $M_{\rm min} = 0.125 + 2.86 \hspace{1 pt} \J^{0.47}$ & Outdated cooling rate,$^{\dagger}$ neglect \\
 
 &&&& self-shielding \\
 
 \\
 
 \citet{Yoshida2003} & $\J = 0-0.01$ & Simulations & $M_{\rm min} \simeq 0.7-0.9$ & Too low resolution$^{\S}$ \\
 
 \\
 
 \citet{Trenti2009} & $\J \geq 0$ & Analytical & $M_{\rm min} = \max \left[1.61 \hspace{1 pt} z_{10}^{-2.074}, 360 \hspace{1 pt} \J^{0.457} z_{10}^{-3.557}  \right]$ & Unrealistic density structure,$^{\bullet}$ \\

& & & &  neglect self-shielding \\

 \\
 \citet{Hummel2012} & $\J = 0$ & Simulations & $M_{\rm min} = 3.26 \hspace{1 pt} z_{10}^{-1.5}$ & Too low resolution$^{\S}$ \\

 \\
 \citet{Visbal2014_H2} & $\J = 0-1$ & Semi-analytical & $M_{\rm min} = 1.05 \hspace{1 pt} z_{10}^{-1.5} (1 + 22.9 \hspace{1 pt} \J^{0.47})$ & Tuned to simulations without \\
 & & & & self-shielding$^*$ \\

 \\
 \citet{Latif2019_cooling} & $J_{21} = 0.1-10^3$ & Simulations & $M_{\rm min} = \exp[5.56 - 4.3 \times \exp(-0.16 \ln J_{21})]$ & Small sample (6 minihaloes) \\
 
 \\
 \citet{Schauer2019, Schauer2021} & $J_{21} = 0-0.1$ & Simulations & $M_{\rm ave} = 1.04 \times 10^{0.9989 \sqrt{J_{21}}}$ & None known \\
 
 & & & $M_{\rm min} = 0.365 \times 10^{1.55 \sqrt{J_{21}}}$ &  \\
 
 \\
 \citet{Kulkarni2021} & $\J = 0-30$ & Simulations & $M_{\rm 50\%} = 0.196 \hspace{1 pt} (1 + \J)^{0.80} z_{21}^{-1.64(1+\J)^{0.36}}$ & None known \\
 
 \\
 \citet{Lupi2021} & $\J > 0$ & Analytical & $M_{\rm min} \sim 0.1-10$ (see reference for fit) & Neglect self-shielding \\
 
 \\
 \citet{Incatasciato2023_LW} & $\J \simeq 0-15$ & Simulations & $M_{\rm min} \sim 2-200$ (no fit provided) & Too low resolution$^{\S}$ \\
 
 \\
 
\hline
\hline
\end{tabular}

\begin{flushleft}$^{\diamond}$: We note that slightly different definitions of $\beta$ exist in the literature. In particular, \citet{Omukai2001_FUV} and \citet{Shang2010} define $\beta \equiv J_{\nu}(h\nu = 12.4 ~ \textrm{eV})/J_{\nu}(h\nu = 13.6 ~ \textrm{eV})$. This would give $\beta \simeq 0.906$ for a $10^5$ K blackbody spectrum, but $\beta \simeq 3.05$ for a $10^4$ K blackbody spectrum.\end{flushleft}

\begin{flushleft}$^{\dagger}$: \citet{Tegmark1997} use the H$_2$-cooling rate from \citet{Hollenbach1979}, whereas \citet{Abel1998_minimummass} and \citet{Machacek2001} use the H$_2$-cooling rate from \citet{Lepp1983}. Both are now known to significantly overestimate the H$_2$-cooling rate \citep[see e.g. figure A1 in][]{Galli1998}.\end{flushleft}

\begin{flushleft} $^{\S}$: \citet{Yoshida2003}, \citet{Hummel2012}, and \citet{Incatasciato2023_LW} use gas particle masses of $m_{\rm gas} = 42.3-142.9 ~ \MSUN$, $m_{\rm gas} = 484 ~ \MSUN$, and $m_{\rm gas} = 1253 ~ \MSUN$ respectively. For comparison, \citet{Greif2011} found that a gas mass resolution of $m_{\rm gas} = 46.4 ~ \MSUN$ increase the halo mass at collapse by a factor $\sim 1.6$ for one of their haloes when compared to their highest resolution runs with $m_{\rm gas} = 0.09-0.72 ~ \MSUN$ \citep[also see][]{Schauer2019}.\end{flushleft}

\begin{flushleft}$^\bullet$: To calculate the cooling rate in a minihalo, \citet{Trenti2009} assume a gas density of $\fB \rho_{\rm vir} \propto (1+z)^3$, i.e. scaling with the virial density. But in reality, as discussed in Section \ref{sec:Halo properties}, the central gas density in minihaloes is determined by adiabatic compression, with different mass and redshift-dependent scalings. \end{flushleft}

\begin{flushleft}$^*$: \citet{Visbal2014_H2} tune their model assumptions to reproduce the halo masses at collapse from \citet{Machacek2001}, \citet{Wise2007_LW}, and \citet{Oshea2008}, none of which take self-shielding into account.\end{flushleft}
\label{TableFitMcrit}
\end{table*}

The purpose of this paper is to present a new detailed analytical calculation of the cooling threshold that can be incorporated into analytical and semi-analytical models of the first stars and galaxies at Cosmic Dawn. Indeed, the results of this paper will be used in one such model in later papers in this series. 

The outline of the paper is as follows. In Sec. \ref{CoolingSection} we calculate the cooling threshold in detail, considering several 
important processes. We start in Sec. \ref{sec:Halo properties} by discussing the gas properties in haloes, which will be crucial for subsequent calculations of the cooling rate. Next, in Sec. \ref{sec:H2threshold}, we analytically derive the H$_2$-cooling threshold in the absence of radiative feedback and streaming velocities and compare the analytical predictions to numerous cosmological simulations. In Sec. \ref{sec:LWfeedback} we generalize this to include LW feedback using semi-analytical methods. We then develop models for photoionizing feedback (Sec. \ref{sec:reionfeedback}), and also consider baryon streaming velocities (Sec. \ref{sec:streaming}). We summarize our calculations in Sec. \ref{sec:coolingsummary} with a final expression for the cooling threshold that incorporates all of the previously mentioned effects. Finally, we conclude in Sec. \ref{SummaryAndConclusion}.

\section{The cooling threshold at Cosmic Dawn}
\label{CoolingSection}

In this section we derive the cooling threshold in haloes at Cosmic Dawn, considering several cooling agents (H$_{2}$, atomic hydrogen, metals), and the effects of radiative feedback. Whenever we need to evaluate expressions numerically in this paper, we adopt the cosmological parameters $\OmegaM = 0.315$, $\OmegaB = 0.0493$, $\OmegaL = 0.685$, and $h = 0.674$, consistent with the latest data from Planck \citep{Planck2018}. The resulting universal baryon fraction is $\fB \equiv \OmegaB/\OmegaM = 0.157$, and we will assume a primordial helium mass fraction of $Y_{\rm He} = 0.247$ \citep[see e.g.][and references therein]{Cooke2018}. 
With these cosmological parameters, the resulting mean matter density at redshift $z$ is
\begin{align}
\rhoM  = \frac{3 \OmegaM H_{0}^2}{ 8 \pi G } (1+z)^3 \simeq 2.69 \times 10^{-27} ~ \Z^3 ~ \rm g ~ cm^{-3} ~ \ , \label{rhoB}
\end{align}
and mean baryon density $\rhoB = \fB \rhoM$.
Assuming primordial gas which is approximately atomic and neutral,\footnote{We are mainly concerned with very low metallicity gas, so the metallicity can be ignored. Furthermore, in the case of minihaloes with gas temperatures $< 10^{4}$ K, the gas is largely neutral. As for the assumption of the gas being largely atomic, this is a good approximation for the low molecular hydrogen fractions in these minihaloes.} the mean molecular weight is $\mu = (X + Y/4)^{-1} = 1.23$, where $X = 1 - Y$ is the hydrogen mass fraction. 
The ratio between the hydrogen number density and the total gas number density is, therefore, $\nH/n = \mu X = 0.926$. These relations will often be used below to evaluate expressions numerically.

\subsection{Halo properties}
\label{sec:Halo properties}

To study the halo mass cooling threshold we will need to know gas properties as a function of the halo mass and redshift. A virialized halo of total (dark matter + baryon) mass $M$ at redshift $z$ is expected to have a virial density $\rhovir = \deltavir \rhoM$, where $\deltavir \simeq 18 \pi^2$ is the expected virial overdensity at the high redshifts of interest \citep[see e.g.][for more general expressions]{Tegmark2006}. The resulting virial radius $\Rvir$, virial velocity $\vvir$, and virial temperature $\Tvir$ of the halo are approximately given by \citep[e.g.][]{Barkana2001}
\begin{align}
\Rvir ~ &= ~ \left( \frac{3M}{4 \pi \rhovir} \right)^{1/3} = ~323 ~ M_{6}^{1/3} \left ( \frac{1 + z}{10} \right )^{-1} ~ \mathrm{pc} ~, \label{Rvir} \\ 
\vvir ~ &= ~ \left( \frac{GM}{\Rvir} \right)^{1/2} = ~  3.65 ~ M_{6}^{1/3} \left ( \frac{1 + z}{10} \right )^{1/2} ~ \mathrm{km ~ s^{-1}} ~, \label{vvir} \\
\Tvir ~ &= ~ 0.75 \times \frac{\mu m_{\rm H} \vvir^2}{2 \kB} = ~ 742 ~ M_{6}^{2/3} \left ( \frac{1 + z}{10} \right ) ~ \mathrm{K} ~,
\label{Tvir}
\end{align}
where $M_{\rm x} \equiv M/10^{\rm x} ~ \MSUN$. As the halo collapses and virializes, the gas temperature --- in the case where cooling is inefficient --- is expected to shock heat to a temperature $\simeq \Tvir$, or to be raised to $\Tvir$ by adiabatic compression. The factor $0.75$ in the expression for $\Tvir$ is a correction factor suggested by \citet{Fernandez2014} to better reproduce results from simulations. 

Cosmological hydrodynamic simulations and simple physical arguments suggest that prior to efficient cooling the gas settles in a cored density profile \citep[e.g.][]{Machacek2001, Wise2007, Visbal2014, Inayoshi2015, Chon2016}. The constant-density core has a radius $R_{\rm core} \simeq 0.1 \Rvir$, with an outer envelope where the gas density falls with radial distance according to $n \propto r^{-2}$ \citep[][]{Visbal2014}. We can therefore approximate the gas number density profile in the halo prior to cooling and collapse as follows:
\begin{equation}
n(r) \simeq \frac{n_{\rm core}}{1 + (r/R_{\rm core})^2} \hspace{1 pt} . \label{nprofile}
\end{equation}
This approximate density profile is only used when we need to integrate radially (as in Sec.~\ref{RtypeFrontsection}), or consider boundary conditions at large radii (as in Sec.~\ref{sec:reionfeedback}). For future reference, Eq. (\ref{nprofile}) predicts that half of the gas mass is contained within a radius $R_{\rm 1/2} \simeq 0.566 \Rvir$ if $R_{\rm core} = 0.1 \Rvir$. The central gas density $n_{\rm core}$ is approximately given by \citep[][]{Visbal2014}
\begin{equation}
n_{\rm core} = \mathrm{min} \left[ \frac{\rhoB}{\mu m_{\rm H}} \left( \frac{\Tvir}{T_{\rm IGM}} \right)^{3/2}, ~ 3.5 \Z^3 ~ \mathrm{cm}^{-3} \right] ~ .
\end{equation}
In massive haloes (large $\Tvir$), the central gas density has the same redshift scaling as the virial density.\footnote{\citet{Visbal2014} use physical arguments to suggest an upper limit to the core density of $n_{\rm core} = 5.3 ~ [(1+z)/10]^3 ~ \rm cm^{-3}$. Using the same argument \citet{Inayoshi2015} find an upper limit of $n_{\rm core} = 5.4 ~ [(1+z)/10]^3 ~ \rm cm^{-3}$. However, judging from the high-resolution cosmological simulations of \citet{Visbal2014} (their Figures 1-3), the central gas density in haloes with masses $M_{6} \sim 1-100$ is typically smaller than this upper limit by a factor $\sim 0.6-0.7$. Thus, we adopt a \textit{typical} core gas density of $n_{\rm core} = 3.5 ~ [(1+z)/10]^3 ~ \rm cm^{-3}$ in these haloes. The \textit{typical} core gas density is of greater interest to us than the upper limit since we want to determine the typical cooling threshold.} In haloes of lower mass the maximum gas density has an upper limit set by adiabatic compression from the IGM until the virial temperature is reached \citep{Tegmark1997, Barkana2001, Visbal2014}. Using the \texttt{RECFAST} code \citep{Seager1999, Seager2000}, \citet{Tseliakhovich2010} found that the IGM temperature at redshifts $z \ll 114$ is approximately given by $T_{\rm IGM} \simeq 2.29 ~ [(1+z)/10]^2 ~ \rm K$ \citep[see also e.g.][]{Liu2019}. Using this relation, we find 
\begin{equation}
n_{\rm core} = \begin{cases} 1.2 ~ M_{6} ~ \Z^{3/2} ~ \text{cm}^{-3} & \text{for }M_{6} \leq 2.92 ~ \Z^{3/2} \\ 3.5 ~ \Z^3 ~ \mathrm{cm}^{-3} & \text{for }M_{6} > 2.92 ~ \Z^{3/2} \end{cases} ~ .
\label{ncore}
\end{equation}
While the above-mentioned simulations of \cite{Visbal2014} and related analytical arguments did not include radiative cooling, we note that \cite{Kulkarni2021} found that the linear scaling $n_{\rm core} \propto M$ with halo mass for minihaloes at fixed redshift to be a good approximation up until the cooling threshold is reached in their simulations which do include radiative cooling (see their fig. 6 and related discussion). 

To ascertain whether cooling is efficient in a given halo, we will need to compare the cooling time-scale $\tcool$ to the free-fall time-scale $\tff$. The free-fall time-scale in the constant-density core (where cooling would be most efficient) is given by $\tff = (3\pi/32 G \bar{\rho}_{\rm core})^{1/2}$, where $\bar{\rho}_{\rm core} \simeq \bar{\rho}_{\rm DM, core} + \mu m_{\rm H} n_{\rm core}$ is the average total matter density within the core (i.e. $r < 0.1 \Rvir$). Assuming a Navarro-Frenk-White \citep[NFW,][]{NFW1995, NFW1996, NFW1997} density profile for the dark matter with a typical halo concentration $c \simeq 3$, we find $\bar{\rho}_{\rm DM, core} \simeq  2.0 \times 10^{-23} ~ [(1+z)/10]^{3} ~ \text{g cm}^{-3}$.\footnote{The average dark matter density in the constant-density gas core is $\bar{\rho}_{\rm DM, core} = 3M_{\rm DM}(< R_{\rm core})/4 \pi R_{\rm core}^3$. For the NFW profile the enclosed dark matter mass is $M_{\rm DM}(< R_{\rm core}) = (1-\fB)M f(c R_{\rm core}/\Rvir)/f(c)$, where $f(x) \equiv \ln(1+x) - x/(1+x)$. For haloes with masses $0.1 < M_{6} < 100$ at redshifts $z > 6$, \citet{Correa2015Concentration} find $c \simeq 3$ (see their Figure 7). With $R_{\rm core}/\Rvir = 0.1$, this yields $M_{\rm DM}(< R_{\rm core})/M \simeq 0.042$, and so $\bar{\rho}_{\rm DM, core}/\rhovir \simeq 42$.} This value is significantly larger than the central gas mass density (especially for the minihaloes of interest), as this gives $\mu m_{\rm H} n_{\rm core}/\bar{\rho}_{\rm DM, core} \lesssim 0.36$ for all haloes. Thus, we can take $\bar{\rho}_{\rm core} \simeq  \bar{\rho}_{\rm DM, core}$, giving us
\begin{equation}
\tff \simeq 1.5 \times 10^{7} ~ \Z^{-3/2} ~ \rm yrs  ~ \label{tff}.
\end{equation}
For future reference, the cooling time-scale in the central region of the halo is $\tcool = (3 n_{\rm core} \kB \Tvir/2) / \Lambda(n_{\rm core}, \Tvir)$, where $\Lambda(n,T)$ is the cooling rate (in $\text{erg s}^{-1}\text{ cm}^{-3}$) at a gas number density $n$ and temperature $T$. Using Eq. (\ref{ncore}), we find
\begin{equation}
\tcool = \begin{cases} ~ t_{\rm cool,LM} & \text{for }M_{6} \leq 2.92 ~ \Z^{3/2} \\   ~ t_{\rm cool,HM} & \text{for }M_{6} > 2.92 ~ \Z^{3/2} \end{cases} ~ ,
\label{tcool}
\end{equation}
where the cooling time-scales in the low-mass ($t_{\rm cool,LM}$) and high-mass ($t_{\rm cool,HM}$) limits respectively are given by
\begin{align}
t_{\rm cool,LM} ~ &= ~ \dfrac{5.8 \times 10^{4} ~ M_{6}^{5/3}}{\Lambda_{-25}(n_{\rm core}, \Tvir)} ~ \Z^{5/2}  ~ \text{yrs} ~ , \label{tcoolLM} \\ 
t_{\rm cool,HM}  ~ &= ~ \dfrac{1.7 \times 10^{5} ~ M_{6}^{2/3}}{\Lambda_{-25}(n_{\rm core}, \Tvir)} ~ \Z^4  ~ \text{yrs} ~ , \label{tcoolHM}
\end{align}
where $\Lambda_{-25}(n,T) \equiv \Lambda(n,T)/10^{-25} ~ \text{erg s}^{-1}\text{ cm}^{-3}$.


\subsection{The H$_{2}$-cooling threshold: No radiative feedback}
\label{sec:H2threshold}

Within minihaloes with virial temperatures $\Tvir < 10^4 ~ \rm K$, the dominant cooling agent is molecular hydrogen, H$_2$. For temperatures $120~\text{K} < T < 6400~\text{K}$, \citet{Trenti2009} found that the H$_2$-cooling rate of \citet{Galli1998} could be approximated as a power law \citep[see also p. 17 in][]{Stiavelli2009}:\footnote{More recent estimates of the H$_2$-cooling rate than provided by \citet{Galli1998} do exist in the literature, but at temperatures, $T > 1000 ~ \rm K$ most relevant to us the differences are small and can be neglected for our purposes. In particular, \citet{Glover2008} find the cooling rate to drop relative to the cooling rate of \citet{Galli1998} at lower temperatures ($T < 1000 ~ \rm K$). However, even more recent calculations by \citet{Coppola2019} and \citet{Flower2021} show \citet{Glover2008} slightly underestimate the cooling rate for $T < 1000 ~ \rm K$ by a factor of $\sim 2$ or so, which render these recently calculated cooling rates closer to \citet{Galli1998}.}
\begin{align}
\Lambda_{\rm H_{2}}(n,T)  ~ &= ~ 3.98 \times 10^{-35} ~ T^{3.4} n_{\rm H}n_{\rm H_{2}} ~ \text{erg cm}^{-3}\text{ s}^{-1} ~ , \\ 
  ~ &= ~ 3.41 \times 10^{-35} ~ T^{3.4} \fmol n^{2} ~ \text{erg cm}^{-3}\text{ s}^{-1} ~ , \label{H2cool}
\end{align}
where $\fmol \equiv n_{\rm H_{2}}/n_{\rm H}$ is the molecular hydrogen fraction. With some foresight, we assume that we are dealing with minihaloes with masses $M_{6} < 2.92 ~ [(1+z)/10]^{3/2}$ (the lower mass range in Eq. \ref{ncore}). We will later see that the cooling threshold indeed falls in this range, justifying this assumption. Using $n_{\rm core}$ from Eq. (\ref{ncore}) and $\Tvir$ from Eq. (\ref{Tvir}) yields 
\begin{equation}
\Lambda_{\rm H_{2},-25}(n_{\rm core},\Tvir) \simeq 2.82 ~ M_{6}^{4.27} \Z^{6.4} \fmol ~,
\end{equation}
which is the H$_2$ cooling rate in units of $10^{-25}$ erg s$^{-1}$ cm$^{-3}$.
Thus, if H$_2$ builds up an abundance $\fmol$ in the central region of a virialized minihalo, the gas will cool on a time-scale (see Eq. \ref{tcoolLM})
\begin{equation}
t_{\rm cool, H_{2}} \simeq 2.1 \times 10^{4} ~ M_{6}^{-2.60} \Z^{-3.9} \fmol^{-1} ~ \text{yrs} ~.
\end{equation}
If $t_{\rm cool, H_{2}} < \tff$, the gas will rapidly lose pressure support and undergo near free-fall collapse to the centre. This scenario is expected to correspond to the \textit{maximum} cooling threshold (call it $M_{\rm max}$), above which the vast majority of haloes can host cool gas. However, it is possible in rare cases that minihaloes with negligible dynamical heating from mergers have much more time for (isobaric) cooling to take place \citep{Yoshida2003, Reed2005_dynheat}. The minimum cooling threshold ($M_{\rm min}$) is, therefore, likely to correspond to the requirement that $t_{\rm cool} < t_{\rm uni}$, where $t_{\rm uni}$ is the age of the Universe. We have $t_{\rm uni}/t_{\rm ff} \simeq 36.3$ independent of redshift.\footnote{In the high-redshift matter-dominated era of interest, the age of the Universe is $t_{\rm uni} \simeq (2/3)H_{0}^{-1} \OmegaM^{-1/2} (1+z)^{-3/2} = 5.45 \times 10^{8} \hspace{1 pt} [(1+z)/10]^{-3/2} ~ \rm yrs$.} Thus, to be general, we can consider the cooling threshold as $t_{\rm cool} < \eta t_{\rm ff}$, with $\eta = 1$ and $\eta = 36.3$ corresponding to $M_{\rm max}$ and $M_{\rm min}$, respectively. The majority of haloes are expected to cool above a mass $M_{50\%}$ that fall somewhere between the two limits, i.e. $M_{\rm min} < M_{50\%} < M_{\rm max}$ (or equivalently, $1 < \eta < 36.3$).

From Eq. (\ref{tff}) we see that $t_{\rm cool} < \eta t_{\rm ff}$ when $\fmol$ exceeds a threshold value:
\begin{equation}
f_{\rm H_{2}, crit} \simeq 1.4 \times 10^{-3} ~ \eta^{-1} M_{6}^{-2.60} \Z^{-2.4} ~. \label{fmolcrit}
\end{equation}
For comparison, \citet{Trenti2009} found $f_{\rm H, crit} \simeq 0.9 \times 10^{-3} ~ M_{6}^{-1.60} [(1+z)/10]^{-3.9}$ (their eq. 7). The different scaling with the halo mass and redshift stems from the fact that we assume --- in line with simulations and physical arguments --- that the central gas density scale as $\propto M (1+z)^{3/2}$ (Eq. \ref{ncore}) rather than $\propto (1+z)^3$ as assumed by \citet{Trenti2009}. 

To find the H$_2$-cooling threshold halo mass, we need to determine the minimum halo mass such that $\fmol > f_{\rm H_{2}, crit}$. In gas of low metallicity and modest density, H$_2$ mainly forms in the following steps \citep[e.g.][]{Pagel1959, McDowell1961, Peebles1968, Dalgarno1973, Tegmark1997}:
\begin{align}
 \text{H} + e^{-} ~ &\longrightarrow ~  \text{H}^{-} + h\nu ~~ (\text{rate }k_{2}) \label{H2form_step1} \\
 \text{H} + \text{H}^{-} ~ &\longrightarrow ~ \text{H}_{2} + e^{-} ~~ (\text{rate }k_{3}) ~. \label{H2form_step2}
\end{align}
The formation of H$^{-}$ is the limiting step in these reactions \citep[see e.g.][]{, Tegmark1997, Glover2006, Glover2007}. One can, therefore, assume that H$^{-}$ reaches its equilibrium abundance, $n_{\rm H^{-}} = k_{2} n_{\rm e}/k_{3}$.\footnote{To see this, we can write down the differential equation for the number density of H$^{-}$: $\Dot{n}_{\rm H^{-}} = k_{2} n_{\rm H^0} n_{\rm e} - k_{3} n_{\rm H^0} n_{\rm H^{-}}$, where $n_{\rm H^0}$ is the number density of neutral hydrogen atoms. Assuming a constant electron number density $n_{\rm e}$, the general solution is $n_{\rm H^{-}} = \mathcal{B} \exp(- k_{3} n_{\rm H^0} t) + k_{2}n_{\rm e}/k_{3}$, where $\mathcal{B}$ is a constant set by the initial conditions. Thus, H$^{-}$ approach its equilibrium abundance $k_{2}n_{\rm e}/k_{3}$ on a time-scale $t_{\rm eq, H^{-}} \simeq (k_{3} n_{\rm H^0})^{-1}$ \citep{Glover2006}. For temperatures $100 ~ \text{K} < T < 10^{4} ~ \text{K}$ the rate coefficient is $k_{3} \sim 2 \times 10^{-9} ~ \text{cm}^3 \text{ s}^{-1}$ to within a factor of $\sim 2$ \citep{Kreckel2010, Glover2015}. Thus, $t_{\rm eq, H^{-}} \sim (20/n_{\rm H^0}) ~ \rm yrs$. Since this is extremely short for relevant gas densities, one can safely assume the equilibrium abundance.} If $k_{1}$ is the recombination coefficient, the evolution of the molecular hydrogen fraction $\fmol$ is governed by \citep{Tegmark1997}:
\begin{align}
    \Dot{x} &\simeq - k_{1} \nH x^2 \label{xdot} \\
    \Dot{f}_{\rm H_2} &\simeq k_{2} \nH x ~ \label{fdot},
\end{align}
where $x \equiv n_{\rm H^{+}}/\nH \simeq n_{\rm e}/\nH$ is the ionization fraction, and it has been assumed that the gas is largely atomic and neutral (more specifically, $x + 2 \fmol \ll 1$). Prior to efficient cooling, the gas density in the virialized minihalo is nearly constant with respect to time. In this case we can, following \citet{Tegmark1997}, first solve Eq. (\ref{xdot}) to get $x(t) = x_{0}/(1 + x_{0} \nH k_{1} t)$, where $x_{0}$ is the initial ionization fraction. Using this result in Eq. (\ref{fdot}) then yields \citep[see Eq. 16 of][]{Tegmark1997}:
\begin{equation}
\fmol(t) \simeq \frac{k_{2}}{k_{1}} \ln \left( 1 + x_{0} \nH k_{1} t \right ) ~ , \label{fmol}
\end{equation}
where we have assumed that the initial H$_{2}$ abundance is negligible.\footnote{The initial value of $\fmol$ in the intergalactic medium is expected to freeze out for $z \lesssim 40$ to $\fmol(z \lesssim 40) \simeq 6 \times 10^{-7}$ \citep[][]{Galli2013}. Since this is much smaller than the critical value $f_{\rm H_{2},crit}$ for efficient cooling (Eq. \ref{fmolcrit}), it follows that the initial abundance can be neglected when estimating the halo mass cooling threshold (where, by definition $\fmol = f_{\rm H_{2},crit}$). }

We evaluate this expression numerically. Since the neutral gas in the minihalo is optically thick to Lyman-continuum (LyC) photons,\footnote{The optical depth of the central gaseous core at the Lyman limit ($h\nu = 13.6 ~ \rm eV$) is $\tau_{0} \simeq 0.926 \hspace{1 pt} n_{\rm core} \sigma_{0} R_{\rm core}$, where $\sigma_{0} \simeq 6 \times 10^{-18} ~ \rm cm^2$ is the photoionization cross-section. For a minihalo with $M_{6} \leq 2.92 [(1+z)/10]^{3/2}$ we find $\tau_{0} \simeq 660 \hspace{1 pt} M_{6}^{1/3} [(1+z)/10]^{1/2}$, which indeed is much greater than unity.} a recombination to the ground state will produce an ionizing photon that is promptly absorbed, yielding no net change in the ionization state. It is therefore appropriate to adopt the Case B (rather than Case A) recombination coefficient, which only considers recombinations to excited states of hydrogen. In contrast, \citet{Tegmark1997} appear to have adopted the Case A recombination coefficient in their estimate of $\fmol(t)$. In particular, they use the value for $k_{1}$ from \citet{Hutchins1976}, which in turn can be traced to the Case A recombination coefficient from \citet{Spitzer1956} (pp. 91-92). The ratio between the value of $k_{1}$ adopted by \citet{Tegmark1997} and the Case B recombination coefficient is $\sim 1.5$ near $T \simeq 1000 ~ \rm K$, and increases to $\sim 2$ at  $T = 10^4 ~ \rm K$ \citep[see e.g. the fit to the Case B recombination coefficient on p. 139 in][]{Draine2011}. This overestimate of $k_{1}$ made its way into the calculation of the cooling threshold by \citet{Trenti2009}, and is therefore inherited in even more recent (semi-)analytical models of Pop III star formation \citep{Crosby2013, Griffen2018, Mebane2018, Mebane2020}.

Here we adopt the Case B recombination coefficient, which in the temperature range $100 ~ \text{K} < T < 10^4 ~ \text{K}$ can be approximated as a power law. For the rate $k_{2}$, we adopt the fit provided by \citet{Galli1998}, which is in good agreement with the fit used by \citet{Tegmark1997} \citep[which they take from][]{Hutchins1976} for temperatures $10 ~ \text{K} < T \lesssim 3000 ~ \text{K}$, but is more accurate at higher temperatures \citep[see fig. 2 and the related discussion in][]{Glover2015}. The rates $k_{1}$ and $k_{2}$ used in this paper are summarized in Table \ref{ReactionRates}. With these reaction rates, the molecular hydrogen fraction after a time $t$ in a minihalo of virial temperature $\Tvir$ is:
\begin{align}
\fmol(t)  ~ &\simeq ~ 6.6 \times 10^{-9} \hspace{1 pt} \Tvir^{1.648} \hspace{1 pt} e^{-\Tvir/16200 } \hspace{1 pt} \ln(1 + x_{0} \nH k_{1} t)~ , \\ 
  ~ &= ~ 3.5 \times 10^{-4} M_{6}^{1.099} \Z^{1.648} e^{-0.0458 \hspace{1 pt} M_{6}^{2/3} [(1+z)/10] } \nonumber \\
  &~~~~~~~~~~~~~~~~~~~~~~~~\times ~ \ln(1 + 0.926 \hspace{1 pt} x_{0} n_{\rm core} k_{1} t) ~ . \label{fmolnumerical}
\end{align}
This predicted relation is nearly a power law, $\fmol \propto M^{1.6}$ at fixed redshift, and halo masses in the minihalo range where the factor in the logarithm is close to unity (see Eq. \ref{logarithmicfactor} below). This is consistent with the recent simulations of \cite{Kulkarni2021} who also find $\fmol \propto M^{1.6}$ at $z = 15$ based on their fig. 6, albeit with a higher normalization compared to our results and some other simulations \citep[e.g.][]{Yoshida2003}.\footnote{From their fig. 6 we see that in the absence of Lyman-Werner feedback and streaming velocities, the molecular hydrogen fraction is nearly a power law (consistent over both and non-cooling and cooling haloes). At $M_5 = 1$ they find $\log_{10} \fmol \simeq -3.75$, which increases to $\log_{10} \fmol \simeq -3$ at $M_5 = 3$ (near the cooling threshold they find at $z = 15$). This gives the quoted power law $\fmol \propto M^{1.6}$.}

\begin{figure*}
\includegraphics[trim={1cm 1cm 1cm 1cm},clip,width=0.9\textwidth]{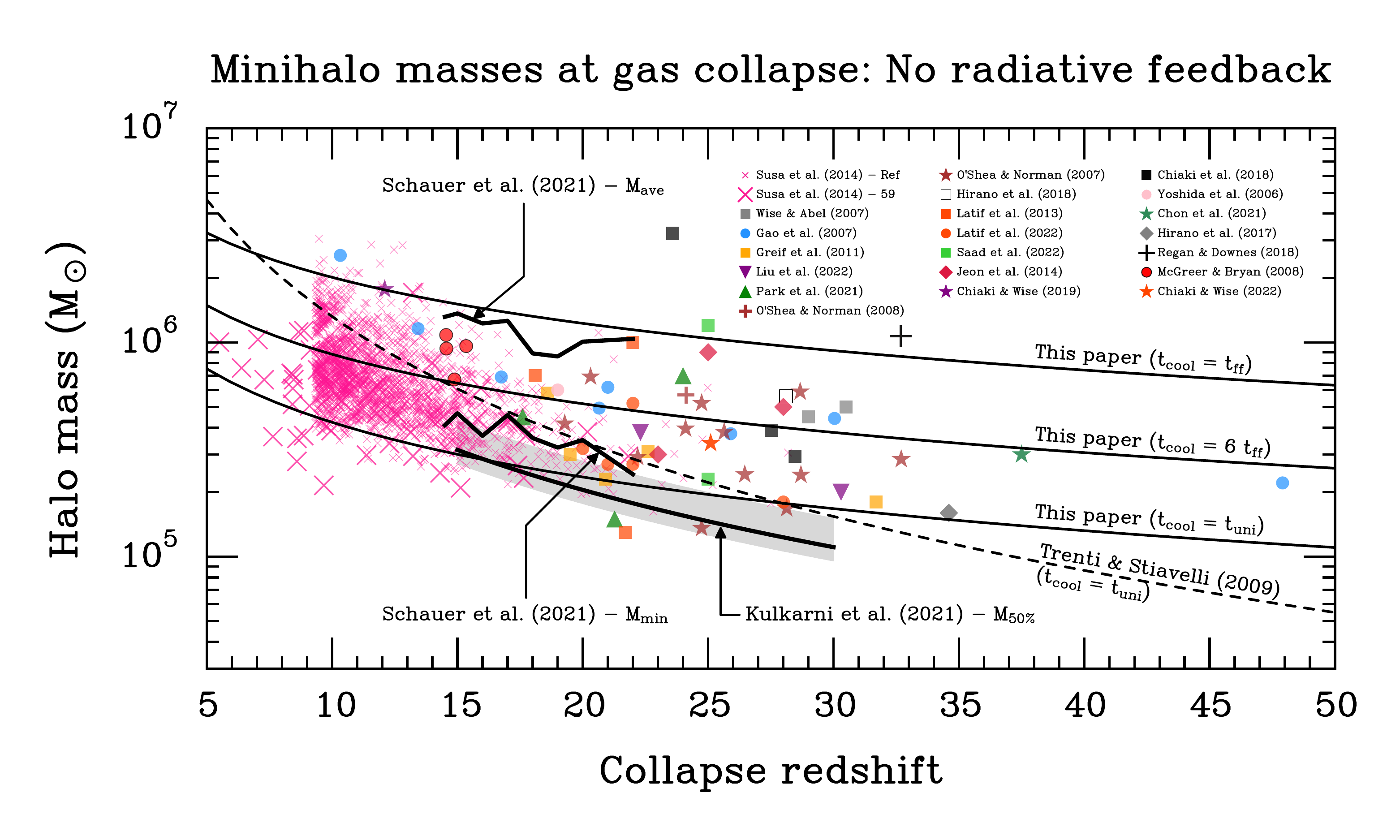}
\caption{A comparison of the cooling thresholds $t_{\rm cool} = \tff$, $t_{\rm cool} = 6 \tff$, and $t_{\rm cool} = t_{\rm uni}$ in case of no radiative feedback (solid lines) with data from cosmological simulations (see Table \ref{TableSims}). The gray band show the mass range where $25\%-75\%$ of haloes contain cool gas in the simulations of \citet{Kulkarni2021}. This can be compared to the minimum ($M_{\rm min}$) and average ($M_{\rm ave}$) mass of haloes containing cool gas from \citet{Schauer2021}, shown as thick solid lines. The dashed line shows the analytical model of \citet{Trenti2009}. It can be seen that the range $\tff < t_{\rm cool} < t_{\rm uni}$ of this paper does a fairly good job of reproducing the scatter in minihalo masses at gas collapse from most simulations. The typical (i.e. median) cooling threshold $M_{50\%}$ from most simulations can be approximately reproduced by the threshold $t_{\rm cool} = 6\tff$. The model of \citet{Trenti2009} on the other hand, which assume $t_{\rm cool} = t_{\rm uni}$, has a significantly stronger redshift dependence than the inferred value of $M_{\rm min}$ from simulations, leading to a clear overestimate of the minimum cooling threshold $M_{\rm min}$ at low $z$. \textbf{Credit:} Raw data from \citet{Schauer2021} and \citet{Susa2014} have been kindly provided to us by Anna T. P. Schauer, and Hajime Susa and Kenji Hasegawa, respectively.
}
\label{H2coolingfigure}
\end{figure*}

\begin{table}
\caption{Hydrogen chemistry: Relevant reaction rates.}
\begin{tabular}{l l}
\hline
\hline
Reaction & Reaction rate (cm$^3$ s$^{-1}$) $^{\dagger}$ \\

\hline
\hline

\vspace{3 pt}

$\text{H}^{+} + e^{-} \longrightarrow \text{H} + h\nu$ & $k_{1} = 2.11 \times 10^{-10} \hspace{1 pt} T^{-0.72}$ \\

\vspace{3 pt}

$\text{H} +  e^{-} ~~\longrightarrow \text{H}^{-} + h\nu$ & $k_{2} = 1.4 \times 10^{-18} \hspace{1 pt} T^{0.928} \hspace{1 pt} e^{-T/16200}$ \\

\vspace{3 pt}

$\text{H}^{-} + \text{H} ~\longrightarrow \text{H}_{2} + e^{-}$ & $k_{3} = \dfrac{c_{1}(T^{c_2} + c_3 T^{c_4} + c_5 T^{c_6})}{1 + c_7 T^{c_8} + c_9 T^{c_{10}} + c_{11} T^{c_{12}}}$ \\

\vspace{3 pt}

$\text{H} + e^{-} ~~\longrightarrow \text{H}^{+} + e^{-}$ & $k_{\rm ion} = 5.86 \times 10^{-11} \hspace{1 pt} T^{0.5} \hspace{1 pt} e^{-157809.1/T}$ \\

\vspace{3 pt}

$\text{H} + \text{H} ~~~\longrightarrow \text{H}_{2}$ & $k_{\rm gr} = 3 \times 10^{-18} \hspace{1 pt} T^{0.5} \left( \dfrac{\mathcal{D}}{\mathcal{D}_{\odot}} \right ) S_{\rm H} \epsilon_{\rm H_2}$ \\

\hline
\hline
\end{tabular}
\vspace{1 pt}\\
$\dagger$ The reaction rate $k_{1}$ is a power law fit to the Case B recombination coefficient in \citet{Draine2011}, and $k_{2}$ comes from \citet{Galli1998}. The rate $k_{3}$ is taken from \citet{Kreckel2010}, with coefficients $c_1 = 1.35 \times 10^{-9}$, $c_2 =  9.8493 \times 10^{-2}$, $c_3 =  3.2852 \times 10^{-1}$, $c_4 =  5.5610 \times 10^{-1}$, $c_5 = 2.7710 \times 10^{-7}$, $c_6 = 2.1826$, $c_7 = 6.1910 \times 10^{-3}$, $c_8 = 1.0461$, $c_9 = 8.9712 \times 10^{-11}$, $c_{10} = 3.0424$, $c_{11} = 3.2576 \times 10^{-14}$, and $c_{12} = 3.7741$. The collisional ionization rate $k_{\rm ion}$ is taken from \citet{Black1981}, and is valid for temperatures $T \lesssim 10^{5} ~ \rm K$ \citep{Black1981, Katz1996}. The H$_2$ formation rate on dust grains $k_{\rm gr}$ is taken from \citet{Hollenbach1979} \citep[also see e.g.][]{Omukai2000, Cazaux2004, Glover2007, FIRE3}. The sticking coefficient $S_{\rm H}$ depend on both the gas temperature $T$ and the dust temperature $T_{\rm gr}$, and is given by $S_{\rm H} = [1 + 0.4(T_2 + T_{\rm gr,2})^{1/2} + 0.2 \hspace{1 pt} T_2 + 0.08 \hspace{1 pt} T_{2}^2]^{-1}$, where $T_2 \equiv T/100 ~ \rm K$ and $T_{\rm gr,2} \equiv T_{\rm gr}/100 ~ \rm K$. The factor $\epsilon_{\rm H_2}$ is the formation efficiency, and $\mathcal{D}/\mathcal{D}_{\odot}$ is the dust-to-gas mass ratio normalized to the Solar value.
\label{ReactionRates}
\end{table}
The initial ionization fraction $x_{0}$ is expected to be close to the residual ionization fraction $x_{\rm res}$ well after recombination ($z \lesssim 50$), i.e. $x_{\rm res} \simeq 1.38 \times 10^{-5} \hspace{1 pt} \OmegaM^{1/2}/(\OmegaB h) = 2.33 \times 10^{-4}$ \citep[e.g. p. 19 in][]{Stiavelli2009}. We see that $\fmol(t)$ exceeds the required H$_{2}$ abundance for efficient cooling (Eq. \ref{fmolcrit}) for halo masses:
\begin{align}
    M_{6} &> \frac{1.45}{\eta^{0.2703}} \frac{e^{0.0124 \hspace{1 pt} M_{6}^{2/3}[(1+z)/10]}}{\ln^{0.2703}(1 + 0.926 \hspace{1 pt} x_{0} n_{\rm core} k_{1} t)} \Z^{-1.094} ~. \label{Eq27}
\end{align}
Next, we need to make a suitable choice for the time $t$ available for H$_{2}$ formation in the minihalo. Since H$_{2}$ formation is significantly more rapid in the dense centre of the virialized minihalo than in the earlier collapse phase, $t$ should be approximately equal to $\eta t_{\rm ff}$, which is the time we allow the gas to cool. We then find
\begin{equation}
    x_{0}n_{\rm core} k_{1} \eta t_{\rm ff} \simeq 0.24 \hspace{1 pt}  \eta M_{6}^{0.52} [(1+z)/10]^{-0.72} ~ .
    \label{logarithmicfactor}
\end{equation}
The logarithmic factor in Eq. (\ref{Eq27}) is therefore of order unity. Because of this, we can solve Eq. (\ref{Eq27}) approximately by iteration. As a zeroth order solution, we can set both the exponential and the logarithm to unity. To get a more accurate solution (the first order iteration), we plug in the zeroth order solution to get:
\begin{align}
    M_{6} > \frac{1.45}{\eta^{0.2703}} \frac{e^{0.016 \hspace{1 pt} \eta^{-0.18} [(1+z)/10]^{0.271}}}{\ln^{0.2703} \left[ 1 + 0.27 \hspace{1 pt} \eta^{0.859} \Z^{-1.29} \right ] } \Z^{-1.094} \label{H2coolingthreshold}.
\end{align}
In Figure \ref{H2coolingfigure} we plot the cooling thresholds $t_{\rm cool} = t_{\rm ff}$ (i.e. $\eta = 1$), $t_{\rm cool} = 6t_{\rm ff}$ ($\eta = 6$), and $t_{\rm cool} = t_{\rm uni}$ ($\eta = 36.3$) from Eq. (\ref{H2coolingthreshold}). Also shown are data from various cosmological simulations, summarized in Table \ref{TableSims}. In particular, we have plotted the typical cooling thresholds ($M_{50\%}$) from \citet{Kulkarni2021} (their eq. 3) and \citet{Schauer2021}. Also shown is the minimum mass of haloes containing cool gas in the simulations of \citet{Schauer2021}. Since there is a discrepancy between the studies of \citet{Schauer2019, Schauer2021} and \citet{Kulkarni2021}, we also plot results from several other cosmological simulations of minihaloes. In all cases, we have chosen minihaloes from simulation runs \textit{without} radiative feedback and baryonic streaming velocities.
\footnote{We have excluded data from some simulations for various reasons. For example, \citet{Machacek2001} and \citet{Hirano2014_PopIII} use old cooling rates not assumed in most other recent simulations, nor in the analytical estimate of this paper. Furthermore, some authors have studied $M_{\rm min}$ but have fairly low resolution: \citet{Yoshida2003} and \citet{Hummel2012} use gas particle masses of $m_{\rm gas} = 42.3-142.9 ~ \MSUN$ and $m_{\rm gas} = 484 ~ \MSUN$, respectively. For comparison, \citet{Schauer2021} have a gas mass resolution of $m_{\rm gas} \simeq 19 ~ \MSUN$ (to within a factor of $\sim 2$), and \citet{Susa2014} have $m_{\rm gas} = 4.90 ~ \MSUN$ for their reference run that contains $1878$ collapsed minihaloes.}

\begin{table}
\caption{Data of minihaloes from cosmological simulations with no radiative feedback or baryon streaming velocities.}
\begin{tabular}{l l l}
\hline
\hline
Reference & Sample & Collapse criteria \\

\hline
\hline

\citet{Schauer2021} & \texttt{v0\_lw0} & $n > 10^2 ~ \rm cm^{-3}$,  \\

 &  & $T < 500 ~ \rm K$, \\

 &  & and $\fmol > 10^{-4}$  \\
 
\\
 
 \citet{Kulkarni2021} & $J_{\rm LW}=v_{\rm bc} = 0$ & $n > 10^2 ~ \rm cm^{-3}$, \\
 
 & & $T < 0.5 \hspace{1 pt} \Tvir$ \\
 
 \\
 
 \citet{Susa2014} & 59 + 1878 & $n > 10^{8} ~ \rm cm^{-3}$ (59), \\
 
 & minihaloes & $n > 10^{3} ~ \rm cm^{-3}$ (1878) \\
 
 \\
 
 \citet{Gao2007} & Tables 1-3 & $n > 10^{10} ~ \rm cm^{-3}$\\
 
 \citet{Wise2007_LW} & \texttt{H2}  & Overdensity $> 10^7$\\
 
 \citet{Greif2011_nostream} & 5 minihaloes & $\nH > 10^9 ~ \rm cm^{-3}$ \\
 
 \citet{Latif2013} & 3 minihaloes & $n \gtrsim 10^{12} \rm cm^{-3}$ \\
 
 \citet{Latif2022} & 5 minihaloes & Sink particle formation \\
 
 \citet{Park2021} & $J_{\rm LW} = J_{\rm X} = 0$ & $\nH > 10^{10} ~ \rm cm^{-3}$ \\
 
 \citet{Oshea2007_12minihaloes} & 12 minihaloes & $n_{\rm H} \gtrsim 10^{10} ~ \rm cm^{-3}$ \\
 
 \citet{Oshea2008} & $J_{\rm LW} = 0$ & $n > 10^{10} ~ \rm cm^{-3}$ \\
 
 \\
 
 \citet{Liu2022} & \texttt{CDM\_A} and & $\nH > 10^{5} ~ \rm cm^{-3}$ \\
 & \texttt{CDM\_B} & \\
 
 \\
 
 \citet{Saad2022} & 2 minihaloes & $n > 10^{12} ~ \rm cm^{-3}$ \\
 
 \citet{Hirano2017_Supersonic} & $J_{\rm LW}=v_{\rm bc} = 0$ & $n \gtrsim 10^{12} ~ \rm cm^{-3}$ \\
 
 \citet{Hirano2018} & \texttt{CDM} & $n > 10^{6} ~ \rm cm^{-3}$ \\ 
 
 \citet{Chiaki2018} & 3 minihaloes & $\nH > 10^{3} ~ \rm cm^{-3}$ \\
 
 \citet{Chiaki2019} & 1 minihalo & $n > 10^{6} ~ \rm cm^{-3}$\\
 
 \citet{Jeon2014} & 3 minihaloes & $\nH > 10^{6} ~ \rm cm^{-3}$\\
 
 \citet{Yoshida2006} & 1 minihalo & $n \gtrsim 3 \times 10^{15} ~ \rm cm^{-3}$\\
 
 \citet{Chon2021} & 1 minihalo & Cold gas presence\\
 
 \citet{Regan2018} & 1 minihalo &  $n \gtrsim \mathrm{few} \times 10^{8} ~ \rm cm^{-3}$ \\
 
 \citet{McGreer2008} & 4 minihaloes & $n > 10^{10} ~ \rm cm^{-3}$ \\
 
 \citet{ChiakiWise2022} & 1 minihalo & $\nH > 10^{6} ~ \rm cm^{-3}$ \\
 
\hline
\hline
\end{tabular}
\label{TableSims}
\end{table}

The largest sample (1937 minihaloes in total) comes from unpublished data from the cosmological simulations of \citet{Susa2014}. This large sample combined with smaller samples from other simulations shows a minimum cooling threshold $M_{\rm min}$ that is more consistent with the result of \citet{Schauer2019,Schauer2021} than \citet{Kulkarni2021}. More specifically, the great majority of minihaloes plotted in Fig. \ref{H2coolingfigure} form cool and dense gas at a mass significantly \textit{above} $M_{75\%}$ from \citet{Kulkarni2021} (upper limit of the gray band in Fig. \ref{H2coolingfigure}), even though only $< 25\%$ of haloes are supposed to undergo gas collapse at this point according to \cite{Kulkarni2021}. 

It is also seen that the cooling threshold $t_{\rm cool} = 6t_{\rm ff}$ of this paper does a fairly good job of reproducing $M_{50\%}$ when considering most of the simulation data. Furthermore, we see that the range $t_{\rm ff} < t_{\rm cool} < t_{\rm uni}$ calculated in this section capture most of the scatter seen in the simulations, consistent with the notion that this range should capture the range of available cooling times in minihaloes. In contrast, the analytical calculation of \citet{Trenti2009}, which assumed $t_{\rm cool} = t_{\rm uni}$, fail to reproduce $M_{\rm min}$ from simulations, with too strong of a redshift dependence. In summary, in the absence of radiative feedback and baryon streaming velocities, we take the typical H$_{2}$-cooling threshold $M_{\rm crit, H_2}$ to be given by Eq. (\ref{H2coolingthreshold}) with $\eta = 6$.

\subsection{The H$_{2}$-cooling threshold: With radiative feedback}
\label{sec:LWfeedback}

As stars form, a significant radiation background builds up. Radiative feedback can either suppress or promote cooling in minihaloes. The most important types of feedback are the following:
\begin{itemize}
\item \textit{Lyman-Werner (LW) feedback}: Photons in the LW band ($11.2-13.6 ~ \rm eV$) can photodissociate H$_{2}$ \citep[e.g.][pp. 346-347]{Draine2011}, and thereby suppress H$_{2}$ cooling in minihaloes.
\item \textit{Photodetachment of H$^{-}$}: The hydrogen anion H$^{-}$ can be photodetached by photons with energies $> 0.754 ~ \rm eV$ \citep[e.g.][]{Miyake2010, McLaughlin2017}, thereby slowing down the formation of H$_{2}$ in minihaloes (see Eqs. \ref{H2form_step1}-\ref{H2form_step2}).
\item \textit{X-ray feedback}: X-ray photons have small photoionization cross-sections and can increase the ionization fraction of the IGM without significantly heating it, resulting in a quicker build-up of H$_{2}$ in minihaloes. This could lower the cooling threshold \citep{Machacek2003_Xrays, Ricotti2016_Xray, Park2021}.
\end{itemize}
Photodissociation of H$_{2}$ and photodetachment of H$^{-}$ are by far the most discussed forms of radiative feedback in minihaloes \citep[e.g.][]{Machacek2001, Wise2007_LW, Wolcott2012_IR, Cen2017_IR, Latif2019_cooling, Skinner2020, Kulkarni2021, Lupi2021, Park2021, Schauer2021} and are the forms of radiative feedback we will consider in this section.

In the presence of LW and IR backgrounds that can photodissociate H$_{2}$ and photodetach H$^{-}$ respectively, Eqs. (\ref{xdot}) and (\ref{fdot}) for the formation of H$_{2}$ change to:
\begin{align}
    \Dot{x} &= k_{\rm ion} n_{\rm H^0} x - k_{1} \nH x^2 \label{xdotLW} \\
    \Dot{f}_{\rm H_2} &= \frac{k_{2} n_{\rm H^0} x}{1 + k_{\rm de}/k_{3} n_{\rm H^0}}  - k_{\rm LW} \fmol ~ \label{fdotLW},
\end{align}
where $n_{\rm H^0} = \nH (1 - x - 2\fmol)$ is the neutral atomic hydrogen number density, and the H$_{2}$ dissociation rate $k_{\rm LW}$ (in units s$^{-1}$) is given by \citep[e.g.][]{Abel1997, Safranek2012_LW, WolcottGreen2017}
\begin{align}
    k_{\rm LW} &= 1.38 \times 10^{-12} \hspace{1 pt}  \label{kdiss} f_{\rm sh} \bar{J}_{\rm LW,21} \hspace{1 pt}, \\
    f_{\rm sh} &= \frac{0.965}{(1+X/b_{5})^\delta} \nonumber \\
    &+ \frac{0.035}{(1+X)^{0.5}} \exp \left[ - 8.5 \times 10^{-4} (1+X)^{0.5} \right ] \hspace{1 pt}, \label{fsh}  \\
    X &\equiv N_{\rm H_{2}}/5 \times 10^{14} ~ \rm cm^{-2} \hspace{1 pt},\\
    b_{5} &\equiv (2 k_{\rm B}T/m_{\rm H_{2}})^{1/2}/10^{5} ~ \textrm{cm s}^{-1} \hspace{1 pt}.
\end{align}
Here $\bar{J}_{\rm LW,21}$ is the frequency-averaged intensity in the LW band in units of $10^{-21}$ erg s$^{-1}$ cm$^{-2}$ Hz$^{-1}$ ster$^{-1}$. The factor $f_{\rm sh}$ takes self-shielding by H$_{2}$ into account (which depends on the H$_{2}$ column density $N_{\rm H_{2}}$). Neutral hydrogen (H\,{\sc i}) can in principle also shield against the LW background due to Lyman lines in the LW band, but this only becomes effective for column densities $N_{\rm HI} \gtrsim \textrm{few} \times 10^{22} ~ \textrm{cm}^{-2}$ \citep[][]{WolcottGreen2011_HIshield,Neyer2022_HIshield}, which exceeds the column densities found in minihaloes before collapse.\footnote{Assuming all the hydrogen is neutral in a halo, the H I column density is $N_{\rm HI} = 0.926 \int_{0}^{\Rvir} \textrm{d}r \hspace{1 pt} n(r) = 0.136 \hspace{1 pt} n_{\rm core} \Rvir$. LW feedback is only effective up to the atomic-cooling threshold $M_{10^4 \rm K} = 5.11 \times 10^7 \hspace{1 pt} [(1+z)/10]^{-3/2} ~ \MSUN$, and so the H I column density before gas collapse for haloes with $\Tvir < 10^4 ~ \rm K$ is $N_{\rm HI} < 1.8 \times 10^{21} \hspace{1 pt} [(1+z)/10]^{3/2} ~  \textrm{cm}^{-2}$.
} 
We, therefore, neglect shielding by \HI and only focus on self-shielding by H$_2$.

The form above for $f_{\rm sh}$ with $\delta = 2$ was first proposed by \citet{Draine1996}. This provides a sufficiently good fit at low densities and temperatures ($T \sim 100 ~ \rm K$), and was recently adopted by \citet{Schauer2021} in the study of the cooling threshold in a LW background. However, at higher temperatures $\delta = 2$ yields inaccurate results when compared to numerical calculations with \MakeUppercase{Cloudy} \citep{Cloudy}. Instead, \citet{WolcottGreen2019} find that the numerical results can be well-fitted with a density and temperature-dependent $\delta$:
\begin{align}
    \delta(n,T) &= A_{1}(T) \hspace{1 pt} e^{-0.2856 \times \log_{10} n } + A_{2}(T) \hspace{1 pt} ,  \label{alpha} \\
    A_{1}(T) &= 0.8711 \times \hspace{1 pt} \log_{10} T - 1.928 \hspace{1 pt} \nonumber, \\
    A_{2}(T) &= - 0.9639 \times \log_{10} T  + 3.892 \hspace{1 pt} \nonumber,
\end{align}
with $n$ given in cm$^{-3}$ and $T$ in K. Here we adopt Eq. (\ref{alpha}), as recently done by \citet{Kulkarni2021} and \citet{Park2021} in their studies of the cooling threshold in minihaloes. 

The expression for the rate $k_{3}$ that we use is listed in Table \ref{ReactionRates}. The photodetachment rate of H$^{-}$ is given by \citep[e.g.][]{Agarwal2015}:
\begin{equation}
    k_{\rm de} = 1.1 \times 10^{-10} ~ \left( \frac{\alpha}{\beta} \right ) \bar{J}_{\rm LW,21} 
\end{equation}
where $\alpha$ and $\beta$ are parameters characterizing the spectrum of photodetaching photons and LW photons, respectively \citep[for details, see][]{Agarwal2015}. For a population of 1 Myr old massive Pop III stars these authors find $\alpha/\beta \simeq 1.8$. For comparison, in the case of a 100 (300) Myr old $Z/Z_{\odot} = 0.05$ Pop II galaxy with a constant star formation rate they obtain $\alpha/\beta \simeq 0.85$ ($\alpha/\beta \simeq 1.2$). Motivated by this estimate, we adopt $\alpha/\beta = 1$ as an approximate value characterizing the cosmological background, built up by continuous star formation over $\sim \textrm{few} \times 100$ Myrs, or a nearby massive Pop II galaxy. We note that some other authors have assumed a $10^{4} ~ \rm K$ blackbody spectrum to model Pop II galaxies \citep[e.g.][]{Shang2010}, which would give $\alpha/\beta \simeq 670$, and hence likely overestimate the feedback from H$^{-}$ photodetachment in minihaloes for a given $\bar{J}_{\rm LW,21}$. On the other hand, some authors model the spectrum of Pop III stars as $10^{5} ~ \rm K$ blackbodies \citep[e.g.][]{Schauer2021}, which would give $\alpha/\beta \simeq 0.1$ and therefore likely underestimate photodetachment feedback from both short-lived Pop III stars and Pop II galaxies.

\begin{figure*}
\includegraphics[trim={1cm 1cm 1cm 1cm},clip,width=0.8\textwidth]{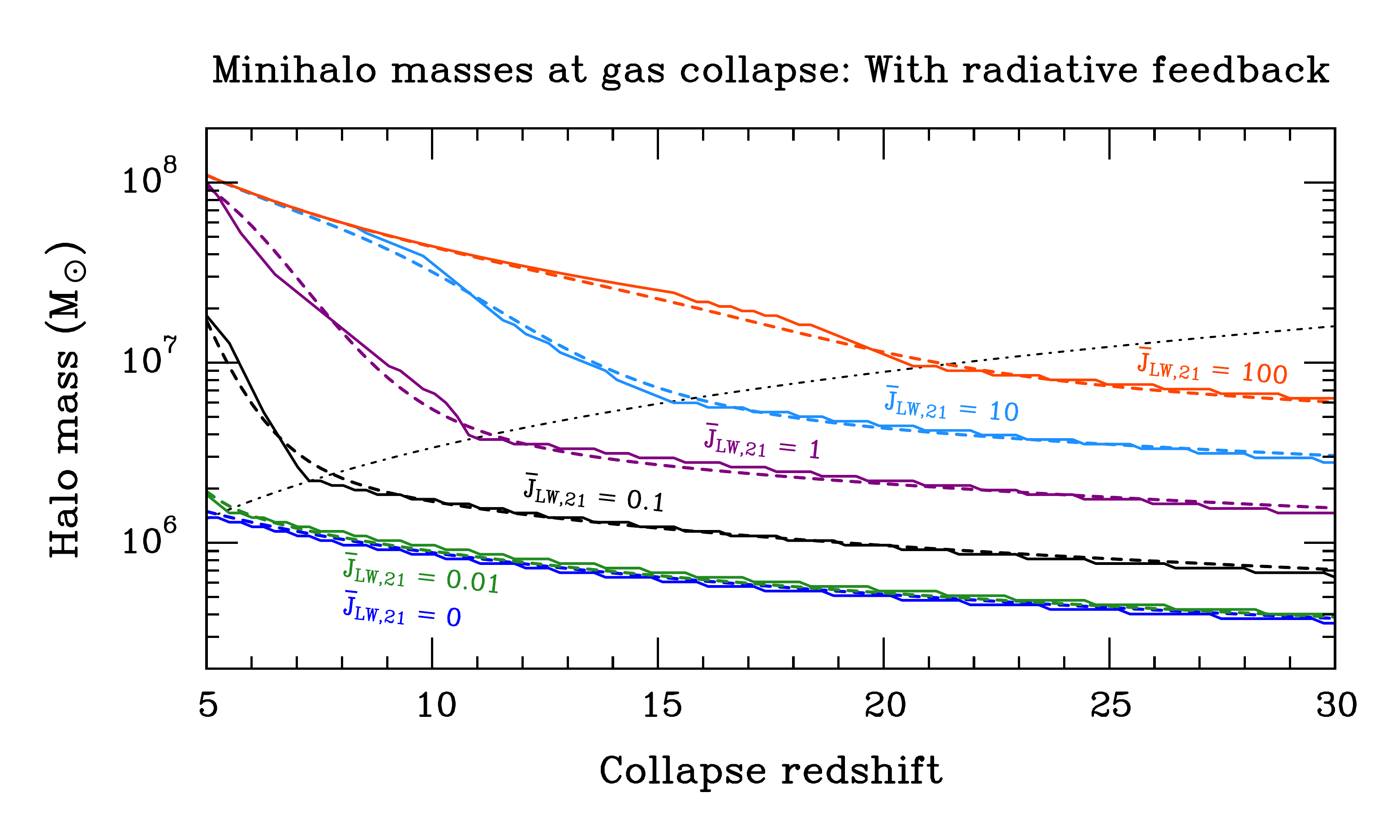}
\caption{The predicted cooling threshold in the presence of a radiation background that can photodissociate H$_2$ and photodetach H$^{-}$. Solid lines are the numerical results, and the dashed lines are the analytical fits to these (Eq. \ref{M_H2LW}). For reference, the dash-dotted line is the halo mass scale $M_{6} = 2.92 \hspace{1 pt} [(1+z)/10]^{3/2}$ above which the central density of the minihalo becomes independent of the halo mass (see Eq. \ref{ncore}), leading to a different redshift evolution of the cooling threshold. Both the numerical results and the fits are colour-coded by the LW intensity $\bar{J}_{\rm LW, 21}$, going from $\bar{J}_{\rm LW, 21} = 0$ (bottom) to $\bar{J}_{\rm LW, 21} = 100$ (top). For $\bar{J}_{\rm LW, 21} \lesssim 0.01$ radiative feedback has a negligible effect on the cooling threshold. }
\label{H2coolingfigureLW}
\end{figure*}

\begin{figure}
    \includegraphics[trim={0.1cm 0.5cm 0cm 0cm},clip,width = \columnwidth]{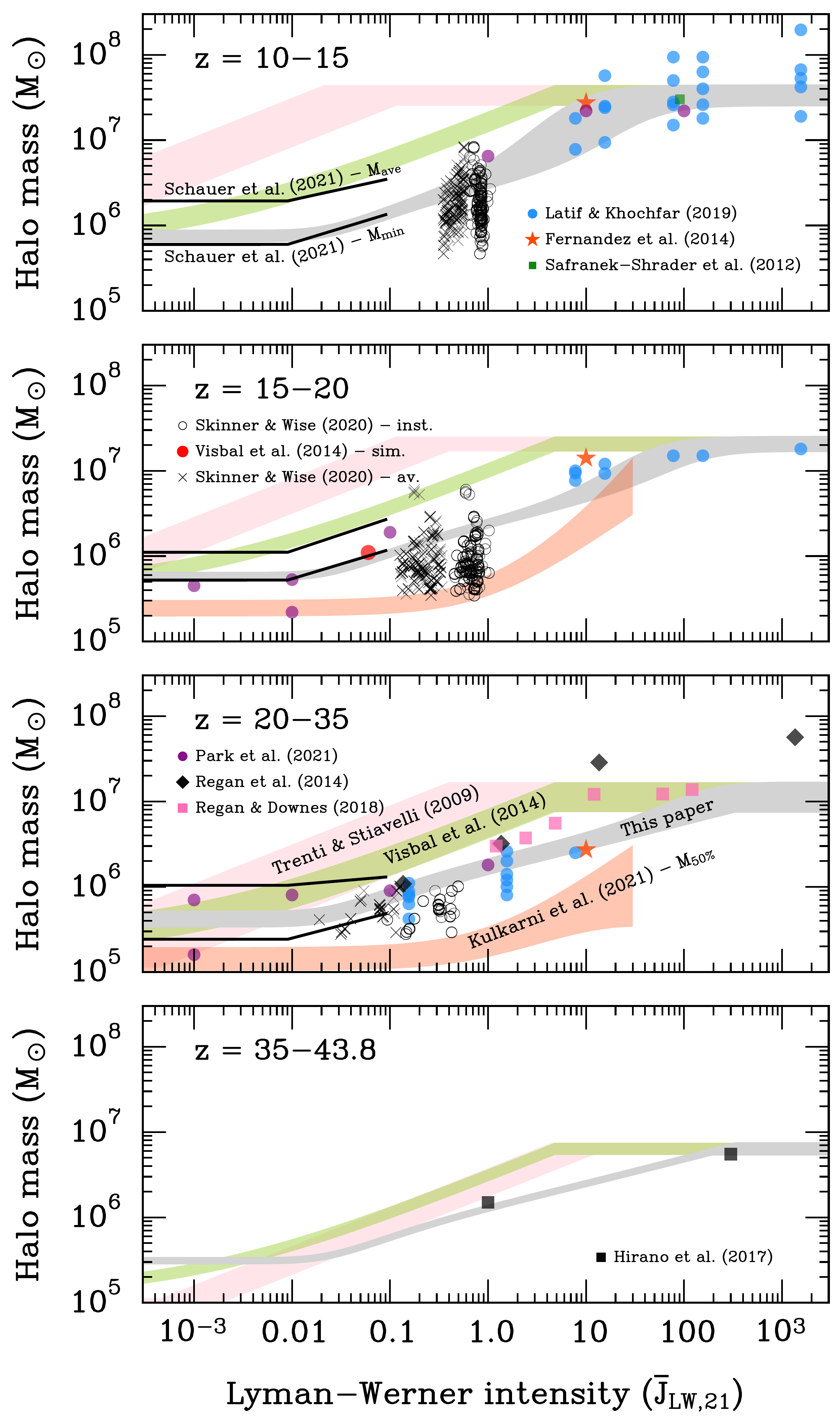}
    \caption{A comparison between our analytically predicted cooling thresholds in a Lyman-Werner (LW) background (grey bands), with those from \citet{Trenti2009} (pink bands), and from \citet{Visbal2014_H2} (green bands). The symbols show minihaloes from cosmological simulations \citep{Safranek2012_LW, Fernandez2014, Regan2014, Visbal2014_H2, Hirano2017_Supersonic, Regan2018, Latif2019_cooling, Skinner2020, Park2021, Schauer2021}. Each subplot shows the predictions and simulation data for a specific redshift interval. Since \citet{Skinner2020} assume a time-dependent LW background we have plotted both their raw data of the local LW background at the time of star formation as black circles, and the cosmic time-averaged LW background as black crosses. Simulation data from \citet{Schauer2021} (black lines) is plotted for $z=14.5$ (in the $z = 10-15$ panel), $z = 18$ ($z = 15-20$ panel), and $z = 22$ ($z = 20-35$ panel). The orange bands in the two middle panels show the results for $M_{50\%}$ from the simulations of \citet{Kulkarni2021} for the redshift range ($z = 15-30$) and LW intensities they consider ($\J \leq 30$). \textbf{Credit:} Raw data from the simulations of \citet{Skinner2020} and \citet{Schauer2021} were kindly provided to us by Danielle Skinner and Anna T. P. Schauer, respectively.
    }
    \label{fig:J21}
\end{figure}

We estimate the cooling threshold (shown in Fig.~\ref{H2coolingfigureLW}) as a function of $\bar{J}_{\rm LW, 21}$ and redshift $z$ numerically as follows:
\begin{enumerate}
\item We consider LW backgrounds $0 \leq \bar{J}_{\rm LW, 21} \leq 100$, redshifts $5 \leq z \leq 30$, and halo masses $3 \times 10^{5} ~ \MSUN \leq M \leq 10^{8} ~ \MSUN$. The total number of cases $N_{\rm LW} \times N_{ z} \times N_{M}$ for which a numerical solution was found is $13 \times 100 \times 100 = 1.3 \times 10^5$.\footnote{The 13 values for the LW backgrounds are $\bar{J}_{\rm LW, 21} = (0, 0.01, 0.1, 1.0, 0.3, 1, 3, 6, 10, 20, 40, 60, 80, 100)$. The 100 redshift points follow a linear and the 100 halo masses a logarithmic distribution over the indicated ranges.} \\
\item For each redshift $z$, halo mass $M$, and LW background $\bar{J}_{\rm LW, 21}$, the temperature evolution of the central region in the halo is estimated by numerically solving the following equation:
\begin{equation}
    \frac{\text{d}T}{\text{d}t} = - \frac{\Lambda_{\rm H_2}(n_{\rm core},T, \fmol)}{3 n_{\rm core} \kB/2 } \hspace{1 pt} ,
\end{equation}
with the initial condition $T(t=0) = \Tvir$. The full expression for the H$_{2}$ cooling rate in low-density gas from \citet{Galli1998} is adopted instead of the power law approximation in Eq. (\ref{H2cool}). The molecular hydrogen fraction and ionization fraction are found simultaneously by solving Eqs. (\ref{xdotLW}) and (\ref{fdotLW}) with the initial conditions $x(t=0) = 2.33 \times 10^{-4}$ and $\fmol(t=0) = 6 \times 10^{-7}$ \citep{Galli2013}. The H$_{2}$ column density of the central region of the halo is estimated as $N_{\rm H_{2}} = 0.926 \hspace{1 pt} \fmol n_{\rm core} R_{\rm core}$, and is fed into Eq. (\ref{fsh}) to compute the self-shielding.\footnote{The column density contribution from the outer envelope of the minihalo, where $\nH \propto r^{-2}$, would only contribute $\Delta N_{\rm H_{2}} \simeq 0.926 \hspace{1 pt } (9/10) \bar{f}_{\rm H_2} n_{\rm core} R_{\rm core}$, where $\bar{f}_{\rm H_2}$ is the radially density-weighted average molecular hydrogen fraction of the outer envelope. Because of the lower density (and hence slower chemical reaction rates) and lower self-shielding in the outer envelope, we expect $\bar{f}_{\rm H_2} \ll \fmol$, where $\fmol$ is the molecular hydrogen fraction in the central core, which is borne out in simulations \citep[see e.g.][]{Oshea2008, Skinner2020}. This justifies our estimate of the H$_2$ column density.}\\

\item The cooling threshold is estimated as the minimum halo mass in the range $3 \times 10^5 ~ \MSUN < M < 10^8 ~ \MSUN$ for which the final temperature is $T(t = 6 \tff) < 0.75 \hspace{1 pt} \Tvir$ after a time $t = 6 \tff$. This criterion for successful cooling is similar to that of \citet{Tegmark1997}, but with the final time being $t = 6 \tff$ instead of a Hubble time, since this was found to better reproduce the typical cooling threshold in the absence of radiative feedback in Sec. \ref{sec:H2threshold}. As we show below (Fig. \ref{fig:J21}), this choice replicates the median results of simulations that incorporate radiative feedback. Redoing the semi-analytical calculation of this section with $t_{\rm cool} < \tff$ and $t_{\rm cool}  < t_{\rm uni}$ would lead to a larger or smaller critical halo mass, respectively -- possibly capturing the scatter seen in simulations, just as in Fig. \ref{H2coolingfigure}. But we are mainly interested in the halo mass threshold above which most haloes can form stars, hence the choice $t_{\rm cool} < 6 \tff$. If no haloes were found to cool down to $T < 0.75 \Tvir$, the cooling threshold is set to $M_{10^4 \rm K} = 5.11 \times 10^7 \hspace{1 pt} [(1+z)/10]^{-3/2} ~ \MSUN$, the atomic-cooling threshold ($\Tvir = 10^4 ~ \rm K$), above which Lyman-$\alpha$ cooling becomes extremely effective \citep[e.g.][]{Oh2002, Fernandez2014, Kimm2016}.
\end{enumerate}

The resulting cooling threshold is plotted in Fig. \ref{H2coolingfigureLW}. These numerical results can be approximately fitted by:
\begin{align}
    M_{\rm crit, H_2 + LW} &= \textrm{min}\left[ M_{\rm crit, H_2} f_{\rm LW} \mathcal{F} + M_{10^4\rm K} (1 - \mathcal{F}), \hspace{1 pt} M_{10^4 \rm K} \right]\hspace{1 pt}, \label{M_H2LW}\\
    f_{\rm LW} &= 1 + 3.5 \hspace{1 pt} \J^{0.31} \hspace{1 pt} e^{- [0.06/\max(\J, 10^{-10})]^{3/4}} \hspace{1 pt} ,\\
    \mathcal{F} &=  \frac{(z/z_{\rm tr})^{7}}{1 + (z/z_{\rm tr})^{7}} \hspace{1 pt}, ~~ z_{\rm tr} = 6.5 \hspace{1 pt} \J^{0.23} \hspace{1 pt} ,
\end{align}
where $M_{\rm crit, H_2}$ is the cooling threshold in the absence of LW feedback, given by Eq. (\ref{H2coolingthreshold}) with $\eta = 6$. Thus, the analytical result in Sec.~\ref{sec:H2threshold} is in good agreement with the more detailed numerical results here for $\J = 0$. As seen in Fig. \ref{H2coolingfigureLW}, the cooling threshold evolves more rapidly with redshift when it reaches $2.92 \times 10^6 \hspace{1 pt} [(1+z)/10]^{3/2} \hspace{1 pt} \MSUN$ (see Fig. \ref{H2coolingfigureLW}), the halo mass above which the central gas density becomes independent of halo mass (Eq. \ref{ncore}). The redshift at which this occurs depends on the value of $\J$, and is captured in the fitting function $\mathcal{F}$ above.

A comparison of Eq. (\ref{M_H2LW}) to cosmological simulations of minihaloes in a LW background is shown in Fig.~\ref{fig:J21}. All the simulation data shown take into account self-shielding, although different authors use different fits in the implementation of this effect. In particular, \citet{Schauer2021} use Eq. (\ref{fsh}) with $\delta = 2$ --- the original fit by \citet{Draine1996}. In contrast to this, \citet{Park2021} and \citet{Kulkarni2021} use the more accurate fit from \citet{WolcottGreen2019}, as done in this paper too. The majority of the simulations plotted in Fig.~\ref{fig:J21} instead adopt the fit recommended by \citet{WolcottGreen2011}, i.e. Eq. (\ref{fsh}) with $\delta = 1.1$ \citep{Safranek2012_LW, Fernandez2014, Regan2014, Visbal2014_H2, Hirano2017_Supersonic, Regan2018, Latif2019_cooling, Skinner2020}. 
\citet{WolcottGreen2019} show that their latest fit to the self-shielding is more accurate than the earlier fits by \citet{Draine1996} and \citet{WolcottGreen2011}. In Appendix \ref{shieldingeffect} we rerun the calculation of the H$_{2}$-cooling threshold with the self-shielding fits from \citet{Draine1996} and \citet{WolcottGreen2011} and show that this can induce errors in the cooling threshold of up to $\sim \textrm{few} \times 10\%$ at the relevant redshifts. 
This error is well within the scatter of simulations, and we are therefore justified in comparing simulations that use these different fits. 

Also shown in Fig.~\ref{fig:J21} are the H$_2$-cooling thresholds in an LW background as derived (semi-)analytically by \citet{Trenti2009} (pink bands) and \citet{Visbal2014_H2} (green bands). For reference, \citet{Trenti2009} did not consider self-shielding. On the other hand, \citet{Visbal2014_H2} performed a similar semi-analytical calculation as in this paper, taking into account self-shielding using the fit from \citet{WolcottGreen2011}. However, they tune their results to simulations that did \textit{not} include self-shielding \citep[][]{Machacek2001, Wise2007_LW, Oshea2008}.

The threshold derived in this paper is consistent with simulations that take self-shielding into account, although significant scatter --- both within and between simulations --- is clearly visible. As in the case of no LW feedback (Fig. \ref{H2coolingfigure}), we see that the simulation results of \cite{Kulkarni2021} for $M_{50\%}$ fall significantly below the results of the other high-resolution simulations, as well as the semi-analytical results of this section. However, like all the other simulations and our calculations, \cite{Kulkarni2021} find that self-shielding drastically reduces the strength of LW feedback.

We also note that \cite{Hirano2015_FUV} simulated Pop III star formation in minihaloes using high-resolution cosmological simulations, including their LW feedback but not a constant LW background. They include self-shielding following \cite{WolcottGreen2011}. A total of 1540 minihaloes form stars in their simulation box. As noted by \cite{Schauer2021}, a direct comparison to their results is complicated by the fact that they do not report the cooling threshold as a function of $\J$, and also because $\J$ is likely fluctuating greatly in time and space within their simulation box. Judging from their figs. 3 and 16, they find a median halo mass at gas collapse of  $M \sim 5 \times 10^5 \hspace{1 pt} \MSUN$ for a median LW intensity $\J \sim \textrm{few} \times 0.01$. These results are broadly consistent with our analytical modelling and most of the simulation results in Fig. \ref{fig:J21}.

In contrast, we find that the analytical formulas derived by \citet{Trenti2009} and \citet{Visbal2014_H2} overestimate the impact of LW feedback when compared to both simulations and our calculations. For example, \citet{Visbal2014_H2} and \citet{Trenti2009} find that $\J = 0.01$ should increase the cooling threshold by a factor $\simeq 3.6$ and $\sim 10$ respectively, whereas simulations and our semi-analytical calculations find a completely negligible increase in the cooling threshold for such an LW background.

\subsection{Metal-cooling haloes}
\label{sec:metalcooling}

When minihaloes are enriched with metals, new avenues for cooling become available. These include:
\begin{itemize}
\item \textit{Cooling via atomic metals like C II and O I}: Both the density and metallicity of gas in primordial haloes prior to collapse is very low, and most of the metals as a result is locked up in either dust grains or atoms, rather than molecules like CO, OH, and H$_2$O \citep[see e.g. discussion in][]{Glover2007}. If cooling via atomic fine-structure lines is efficient enough, then this could potentially trigger collapse and star formation in haloes with $\Tvir < 10^4 ~ \rm K$ \citep[e.g.][]{Bromm2001_metal, Wise2014_metalcooling}.
\item \textit{Cooling via grain-catalyzed H$_{2}$}: In metal-enriched gas, H$_2$ can form on dust grains. At sufficiently high metallicities this can increase the H$_2$ formation rate substantially, and thus lower the H$_2$-cooling-threshold in minihaloes \citep{Nakatani2020}. 
\end{itemize}
Below we discuss and \textbf{estimate} the corresponding cooling thresholds.

\subsubsection{Atomic metal cooling}
\label{CIIcooling}

\begin{figure}
    \includegraphics[trim={0.1cm 0.6cm 0.4cm 0.4cm},clip,width = \columnwidth]{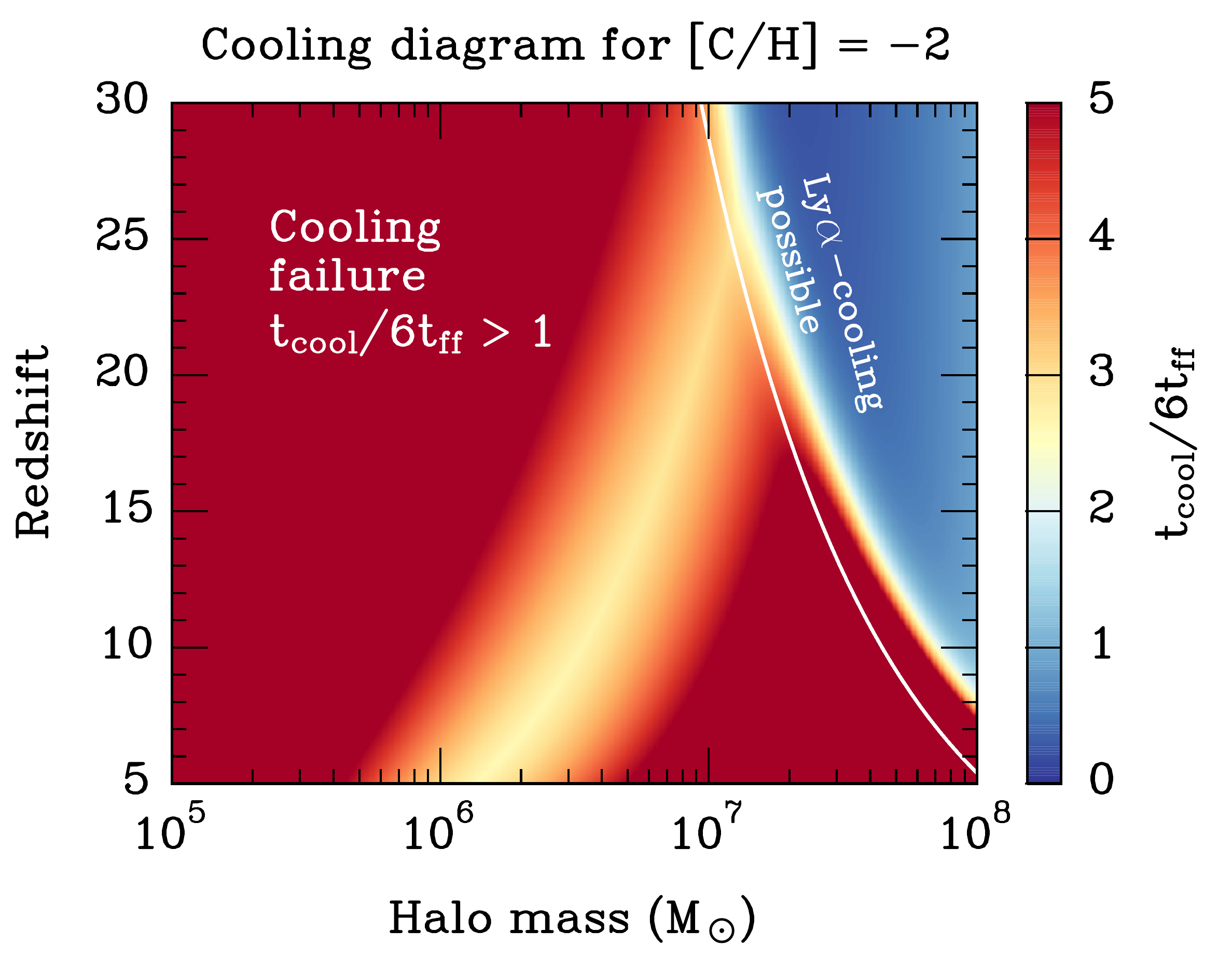}
    \caption{A cooling diagram showing the ratio $t_{\rm cool}/6 \tff$ as a function of halo mass and redshift, assuming $[\textrm{C/H}]=-2$. The white line shows the Ly$\alpha$-cooling threshold $\Tvir = 10^4 ~ \rm K$ for reference, although only the cooling rate from Eq. (\ref{MetalCooling}) was used to make this plot.}
    \label{fig:coolingdiagram}
\end{figure}

At low gas densities,\footnote{More specifically, below the critical densities. The critical hydrogen and electron densities for the [C II]-157.74 $\mu$m line are $n_{\rm crit,H} \simeq 3.2 \times 10^3 \hspace{1 pt} T_{2}^{-0.1281-0.0087 \ln T_2} ~ \textrm{cm}^{-3}$ and $n_{\rm crit,e} \simeq 53 \hspace{1 pt} T_{4}^{1/2}~ \textrm{cm}^{-3}$, respectively \citep[e.g.][p. 197]{Draine2011}. These densities, especially $n_{\rm crit,H}$, are higher than the ones encountered in virialized low-mass haloes prior to gas collapse.} the cooling rate due to atomic metals is approximately given by \citep{FIRE3}:
\begin{align}
    \Lambda_{Z}(n,T) &= 10^{-27} \hspace{1 pt} z_{\rm C} \Big\{ \left( 0.47 \hspace{1 pt} T^{0.15} + 4890 \hspace{1 pt} x T^{-0.5} \right ) e^{-91.211/T} \nonumber \\
    &+ 0.0208 \hspace{1 pt} e^{-23.6/T} \Big\} n_{\rm H}^2 \hspace{1 pt} \label{MetalCooling},
\end{align} 
where $z_{\rm C} \equiv 10^{[\textrm{C/H}]}$, and $x$ the ionization fraction. The dominant term proportional to $e^{-91.211/T}$ in Eq. (\ref{MetalCooling}) comes from the [C II]-157.74 $\mu$m line, while the term proportional to $e^{-23.6/T}$ is due to the [C I]-609.7 $\mu$m line \citep[e.g.][]{Hollenbach1989, Draine2011, Hosuk2016}. In Fig.~\ref{fig:coolingdiagram} we plot the ratio $t_{\rm cool}/6 \tff$ as a function of halo mass $M$ and redshift $z$, assuming a gas temperature $T = \Tvir$, gas density $n = n_{\rm core}$, and carbon abundance $[\textrm{C/H}]=-2$. The electron fraction is taken to be $x = \textrm{max}(2.33\times 10^{-4}, \hspace{1 pt} x_{\rm eq})$, where $x_{\rm eq}$ is the ionization fraction in collisional ionization equilibrium. This prescription maximizes the efficiency of metal cooling in minihaloes. 

We see in Fig.~\ref{fig:coolingdiagram} that below the Ly$\alpha$-cooling threshold ($\Tvir = 10^4 ~ \rm K$) we have $t_{\rm cool}/6\tff \gtrsim 2$. Thus, we only expect a minority of pre-enriched minihaloes with $[\textrm{C/H}]=-2$ to be able to cool via metal fine-structure lines --- namely those rare minihaloes that can cool undisturbed during 
a Hubble time. For minihaloes with significantly lower gas metallicities, we do not expect metal fine-structure cooling to lead to a collapse at all. For this reason, we ignore the effect of metal fine-structure cooling on the critical halo mass for cooling.

\subsubsection{The H$_2$-cooling threshold with dust-catalyzed formation}

\begin{figure}
    \includegraphics[trim={0.1cm 0.4cm 0.2cm 0.4cm},clip,width = \columnwidth]{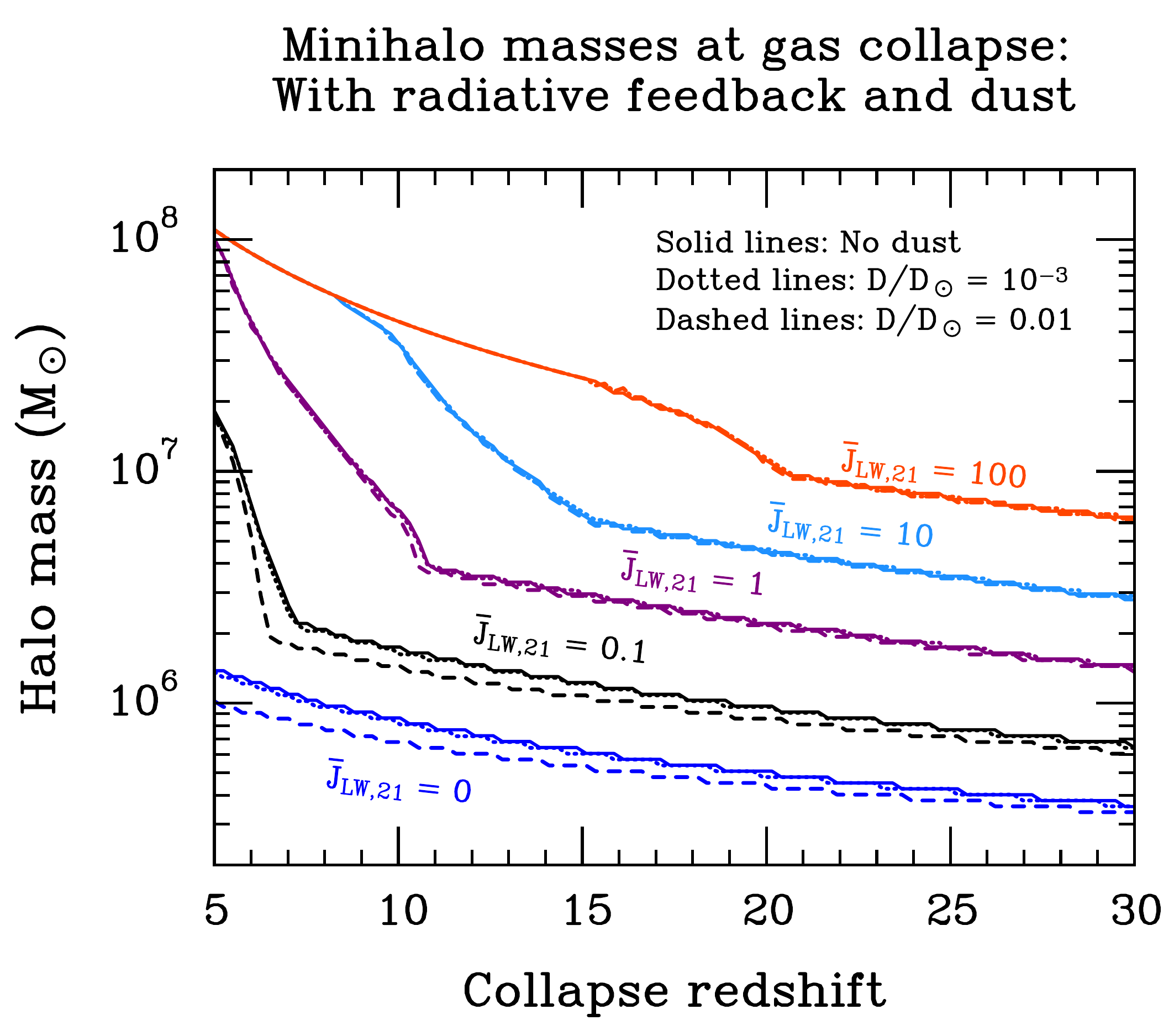}
    \caption{The effect of dust-catalyzed H$_2$ formation on the cooling threshold for different Lyman-Werner (LW) backgrounds. Solid lines reproduce the results without any dust from Figure \ref{H2coolingfigureLW}. The dotted and dashed lines assume $\mathcal{D}/\mathcal{D}_{\odot} = 10^{-3}$ and $\mathcal{D}/\mathcal{D}_{\odot} = 0.01$, respectively.}
    \label{fig:dustcooling}
\end{figure}

A potentially more efficient way of triggering a collapse in metal-enriched minihaloes was suggested by \citet{Nakatani2020}. These authors conducted idealized simulations of minihaloes of different metallicities and 
UV backgrounds, incorporating both molecular hydrogen cooling (with self-shielding) and cooling from metal fine-structure lines. They found that the prime effect of metals on the cooling threshold is the catalyzed H$_2$ formation rate on dust grains \citep[][]{Cazaux2004}. The formation rate on dust grains ($k_{\rm gr}$) is given in Table \ref{ReactionRates}. 
The resulting ratio between the grain-catalyzed formation rate ($\Dot{n}_{\rm H_2, dust} = k_{\rm gr} \nH n_{\rm H^0}$) and the gas phase formation rate ($\Dot{n}_{\rm H_2, gas} = k_{2} n_{\rm H^0} n_{\rm e}$) becomes
\begin{align}
    \frac{\Dot{n}_{\rm H_2, dust}}{\Dot{n}_{\rm H_2, gas}} ~ &= ~ \frac{k_{\rm gr}}{k_2 x} \nonumber \\
    ~ &= ~ 92 \hspace{1 pt} \left( \frac{\mathcal{D}/\mathcal{D}_{\odot}}{0.01} \right ) \left( \frac{x}{2.33\times 10^{-4}} \right )^{-1} T^{-0.428} e^{T/16200} \nonumber \\
    ~ &\times ~ \frac{\epsilon_{\rm H_2}}{[1 + 0.04 (T + T_{\rm gr})^{1/2} + 0.002 T + 8 \times 10^{-6} T^2]} \hspace{1 pt} \label{dust1},
\end{align}{equation}
where $\mathcal{D}/\mathcal{D}_{\odot}$ is the dust-to-gas mass ratio normalized to the solar value, $T_{\rm gr}$ is the dust grain temperature, and $\epsilon_{\rm H_2}$ the H$_2$ formation efficiency on dust grains. In the temperature range $200 ~ \textrm{K} \lesssim T \lesssim 10^4 ~ \textrm{K}$ where H$_2$-cooling is efficient (and probably dominant), we find
\begin{equation}
     \frac{\Dot{n}_{\rm H_2, dust}}{\Dot{n}_{\rm H_2, gas}} \lesssim 4.2 \hspace{1 pt} \epsilon_{\rm H_2} \left ( \frac{\mathcal{D}/\mathcal{D}_{\odot}}{0.01} \right ) \left( \frac{x}{2.33\times 10^{-4}} \right )^{-1} \left( \frac{T}{200 ~ \rm K} \right)^{-0.428} \hspace{1 pt} \label{dust2}.
\end{equation}
Thus, it is conceivable that dust-catalyzed H$_2$ formation could boost cooling significantly in low-mass haloes with $\mathcal{D}/\mathcal{D}_{\odot} > \textrm{few} \times 10^{-3} - 0.01$ if $\epsilon_{\rm H_2} \simeq 1$. 

It is straightforward to include H$_2$ formation on dust in the semi-analytical calculation of the cooling threshold in Sec. \ref{sec:LWfeedback}. The resulting cooling threshold is shown in Fig. \ref{fig:dustcooling} for $\mathcal{D}/\mathcal{D}_{\odot} = 10^{-3}$ and $\mathcal{D}/\mathcal{D}_{\odot} = 0.01$, assuming $\epsilon_{\rm H_2} = 1$ and a dust grain temperature equal to the CMB temperature, $T_{\rm gr} = 2.726 \hspace{1 pt} (1+z) ~ \rm K$. We see that $\mathcal{D}/\mathcal{D}_{\odot} = 10^{-3}$ has no visible impact on the cooling threshold, whereas $\mathcal{D}/\mathcal{D}_{\odot} = 0.01$ decreases the cooling threshold by a $\sim \textrm{few} \times 10\%$ at most, for low redshifts and weak LW backgrounds. For more realistic moderate LW backgrounds ($\J \gtrsim 1$) the effect is weaker still since the cooling threshold is pushed up to higher virial temperatures for which the relative importance of dust becomes negligible (see Eqs. \ref{dust1} and \ref{dust2}). Furthermore, the effect of dust-catalyzed H$_2$ formation may be even weaker for the following reasons:
\begin{enumerate}
\item Dust grains can at most cool down to the CMB temperature $T_{\rm CMB} = 27.26 \hspace{1 pt} [(1+z)/10] ~ \rm K$. The H$_2$ formation efficiency $\epsilon_{\rm H_2}$ is a function of the dust grain temperature $T_{\rm gr}$ \citep[e.g.][]{Hollenbach1979, Cazaux2002, Cazaux2004_efficiency, Cazaux2009, Cazaux2010}. The exact temperature dependence depends on the composition of the grains and their surface properties and is not known precisely. For example, if the barrier width between physisorbed and chemisorbed sites on grains is $2 ~  \textrm{Å}$ ($3 ~ \textrm{Å}$) \citet{Cazaux2010} find $\epsilon_{\rm H_2} < 0.01$ ($\simeq 0.2$) for silicate (olivine) grains, and $\epsilon_{\rm H_2} < 0.01$ ($\simeq 0.5-1$) for carbonaceous grains when the grain temperature is $T_{\rm gr} = 30-100 ~ \rm K$. In this case, dust-catalyzed H$_2$ formation would be even less important in minihaloes at high redshifts. 

\item It is often assumed that the dust-to-gas mass ratio scale linearly with gas metallicity $Z$, so that $\mathcal{D}/\mathcal{D}_{\odot} = Z/Z_{\odot}$. However, there is observational evidence that $\mathcal{D}/\mathcal{D}_{\odot} < Z/Z_{\odot}$ for low metallicities \citep[e.g.][]{Remy2014, DeVir2019}. For example, the results from \citet{Remy2014} could imply that $Z/Z_{\odot} = 0.01$ --- a typical metallicity of an Ultra-Faint Dwarf galaxy or an old globular cluster --- would correspond to $\mathcal{D}/\mathcal{D}_{\odot} \sim 2 \times 10^{-5}$ \citep[][]{Remy2014, Sharda2022}, which makes dust-catalyzed H$_2$ formation completely negligible. 
\end{enumerate}
In summary, we have found both the direct and indirect effects of metals (cooling via metals and H$_2$ formation on grains, respectively) on cooling in low-mass haloes to be negligible at low metallicities of interest \citep[see e.g.][for a similar conclusion]{Jeon2017_UFD}. For our purposes, we, therefore, ignore the effect of metals on the cooling threshold.  

\subsection{Reionization feedback}
\label{sec:reionfeedback}

A strong UV background can not only photodissociate H$_2$ but can, if it extends into the Lyman continuum ($h\nu>13.6$ eV), also photoionize gas, heating it to $T \simeq 2 \times 10^{4} ~ \rm K$. This can prevent gas from collapsing in haloes which have virial temperatures similar to this \citep[e.g.][]{Efstathiou1992, Thoul1996, BarkanaLoeb1999, Kitayama2000, Dijkstra2004, Okamoto2008, Benitez2020}. Indeed, reionization of the IGM is believed to have quenched further star formation in Ultra-Faint Dwarf galaxies \citep[e.g.][]{Bovill2009, Brown2014, Ricotti2016, Fitts2017, Simon2019_UFD, Wheeler2019, Simon2021}. In this section, we analyze the effect of photoionization feedback on the minimum halo mass wherein gas can condense. 

To obtain slightly better accuracy than order-of-magnitude estimates, and to lay the mathematical foundation for subsequent estimates of the central gas accretion rate in the next paper in this series, we start with the fluid equations. Assuming spherical symmetry and an isothermal gas, they read:
\begin{align}
\frac{\partial \Mgas}{\partial t} + v\frac{\partial \Mgas}{\partial r} ~ &= ~ 0 ~ , \\ 
\frac{\partial \Mgas}{\partial r} ~ &= ~  4 \pi r^2 \rhogas ~ , \\ 
\frac{\partial v}{\partial t} + v\frac{\partial v}{\partial r} ~ &= ~ - \frac{\cs^2}{\rhogas} \frac{\partial \rhogas}{\partial r} - \frac{G \Mgas}{r^2} - \frac{G M_{\rm DM}}{r^2}~ , 
\label{Fluidequations}
\end{align}
where $c_{\rm s,h} = (\kB T_{\rm h}/\mu_{\rm h} m_{\rm H})^{1/2}$ is the isothermal sound speed of gas in the halo, $\rhogas(r,t)$ is the gas density, $v(r,t)$ the fluid velocity, $\Mgas = \Mgas(<r,t)$ is the total gas mass enclosed within a radius $r$, and $M_{\rm DM} = M_{\rm DM}(<r,t)$ the total enclosed DM mass within a radius $r$. At this point a self-similarity Ansatz can be made following earlier work on collapsing DM-free isothermal gas clouds \citep[e.g.][]{Shu1977}:
\begin{equation}
v(r,t) = c_{\rm s,h}\Bar{v}(x) \hspace{1 pt}, ~ \Mgas(r,t) = \frac{c_{\rm s,h}^3 t}{G} \Bar{m}(x) \hspace{1 pt}, ~ \rhogas(r,t) = \frac{\Bar{\rho}(x)}{4 \pi G t^2} \hspace{1 pt},
\end{equation}
where $x \equiv r/c_{\rm s,h}t$ (not to be confused with the electron fraction in earlier sections). With these substitutions, and approximating the DM halo as an isothermal sphere using $GM_{\rm DM}(<r)/r^2 = (1 - \fB) v_{\rm vir}^{2}/r$,\footnote{This follows since $M_{\rm DM}(<r) = (1 - \fB) M r/\Rvir$ for an isothermal sphere.} the fluid equations simplify to three ordinary differential equations:
\begin{align}
\frac{\mathrm{d} \Bar{m}}{\mathrm{d} x} (x - \Bar{v}) ~ &= ~ \Bar{m} ~ , \label{Eq11} \\ 
\frac{\mathrm{d} \Bar{m}}{\mathrm{d} x} ~ &= ~  x^2 \Bar{\rho} ~ , \label{Eq12}\\ 
\frac{\mathrm{d} \Bar{v}}{\mathrm{d} x} \left [ (x - \Bar{v})^2 - 1 \right ]  ~ &= ~ \beta \frac{(x - \Bar{v})}{x} + \frac{\Bar{m}}{x^2}(x - \Bar{v})~ , \label{Eq13}
\end{align}
where $\beta \equiv (1 - \fB)(\vvir/c_{\rm s,h})^2 - 2$ is a constant parameter. In the absence of a DM halo we have $\beta = -2$, and we recover the differential equations derived by \citet{Shu1977}.  

We see that the Eqs. (\ref{Eq11}) and (\ref{Eq12}) can be combined to yield $\Bar{m} = (x - \Bar{v}) x^2 \Bar{\rho}$, so that we are free to focus on only the two variables $\Bar{m}$ and $\Bar{v}$. Boundary conditions can be imposed at $x \rightarrow \infty$ for consistency with the assumed initial density profile at large radii, given in Eq. (\ref{nprofile}). To do so, we first determine the enclosed gas mass at large radii:
\begin{align}
    M_{\rm gas}(<r) &= \int_{0}^{r} \textrm{d}r' \hspace{1 pt} 4 \pi r'^2 \mu m_{\rm H} n(r') \\
    &\simeq 4 \pi R_{\rm core}^2 r \mu m_{\rm H} n_{\rm core} \hspace{1 pt} ,
\end{align}
where the second line holds for $r \gg R_{\rm core} \simeq 0.1 \hspace{1 pt} \Rvir$. We therefore find that $\bar{m}(x \rightarrow \infty) \equiv m_{\infty} x$ where
\begin{align}
    m_{\infty} &= \frac{4 \pi G R_{\rm core}^2 \mu m_{\rm H} n_{\rm core}}{c_{\rm s,h}^2} \\
    &\simeq \frac{4 \pi G \Rvir^2 \mu m_{\rm H} n_{\rm core}}{100 c_{\rm s,h}^2} \hspace{1 pt} .
\end{align}
For haloes with $M_6 > 2.92 ~ [(1+z)/10]^{3/2}$ we have $\mu m_{\rm H} n_{\rm core} = 7.2 \times 10^{-24} ~ [(1+z)/10]^{3} ~ \textrm{g cm}^{-3}$, or equivalently $\mu m_{\rm H} n_{\rm core} \simeq 15 \rhovir$. Using this and $G \Rvir^2 \rhovir = 3 \vvir^2 /4 \pi$ yields
\begin{align}
    m_{\infty} &\simeq \frac{45}{100} \frac{\vvir^2}{c_{\rm s,h}^2} \\
    &= 0.53 (\beta + 2) \hspace{1 pt} . \label{minf_largeM}
\end{align}
Let us now consider the gas velocity for $x \rightarrow \infty$. Assuming that $x-\bar{v} \simeq x$ for large $x$ (as expected if the gas starts out stationary at large radii), Eq. (\ref{Eq13}) yields
\begin{align}
    \frac{\mathrm{d} \Bar{v}}{\mathrm{d} x}\Big\vert_{x \rightarrow \infty} &\simeq  \frac{\beta}{x^2} + \frac{\Bar{m}(x \rightarrow \infty)}{x^3} 
    = \frac{\beta + m_{\infty}}{x^2} \hspace{1 pt}.
\end{align}
Upon integrating this expression and imposing the boundary condition $\Bar{v}(x \rightarrow \infty) \rightarrow 0$, we find the following expression for large $x$:
\begin{equation}
    \Bar{v}(x \rightarrow \infty) = - \frac{\beta + m_{\infty}}{x} \hspace{1 pt} . \label{vbar_largex}
\end{equation}
Using Eq. (\ref{minf_largeM}) for $m_{\infty}$ then gives the result that gas accretion onto the halo, which requires $v_{\rm gas} < 0$ at large radii, is only possible for $\beta > -0.693$, or equivalently, $v_{\rm vir} > 1.25 \hspace{1 pt} c_{\rm s, h}$. Reionization is expected to heat the gas to $T_{\rm h} \simeq 2 \times 10^4 ~ \rm K$ \citep[e.g.][]{Okamoto2008, HaardtMadau2012}. If the gas contains fully ionized hydrogen and singly ionized helium the molecular weight is $\mu_{\rm h} = 0.614$ and we find that haloes with virial velocities $\vvir > 21 ~ \textrm{km s}^{-1}$ can retain photoheated gas following reionization. This is consistent with recent cosmological simulations that find that haloes with $\vvir \lesssim 20-30 ~ \textrm{km s}^{-1}$ are quenched of gas by reionization \citep[e.g.][]{Zhu2016, Fitts2017, Jeon2017_UFD, Graus2019, Gutcke2022}. From Eq. (\ref{vvir}) a virial velocity $\vvir = 21 ~ \textrm{km s}^{-1}$ correspond to a halo mass threshold of:
\begin{equation}
    M_{\rm crit, bound} = 1.9 \times 10^{8} ~ \left( \frac{1+z}{10} \right )^{-3/2} ~ \MSUN \hspace{1 pt} . \label{reion_noshield}
\end{equation}
This result is also consistent with the analytical calculation of \citet{Benitez2020}, who modelled gas in hydrostatic equilibrium within NFW haloes in a photoionizing background. However, the result in Eq. (\ref{reion_noshield}) assumes that the gas is not self-shielded against the UV background. For a given photoionizing background, if the gas is sufficiently dense the central region can self-shield and potentially allow cooling and collapse in haloes with masses below $M_{\rm crit, bound}$ \citep[e.g.][]{Kitayama2000, Visbal2017, Kulkarni2019, Nakatani2020}.
\begin{figure}
\centering
    \includegraphics[width = 0.7\columnwidth]{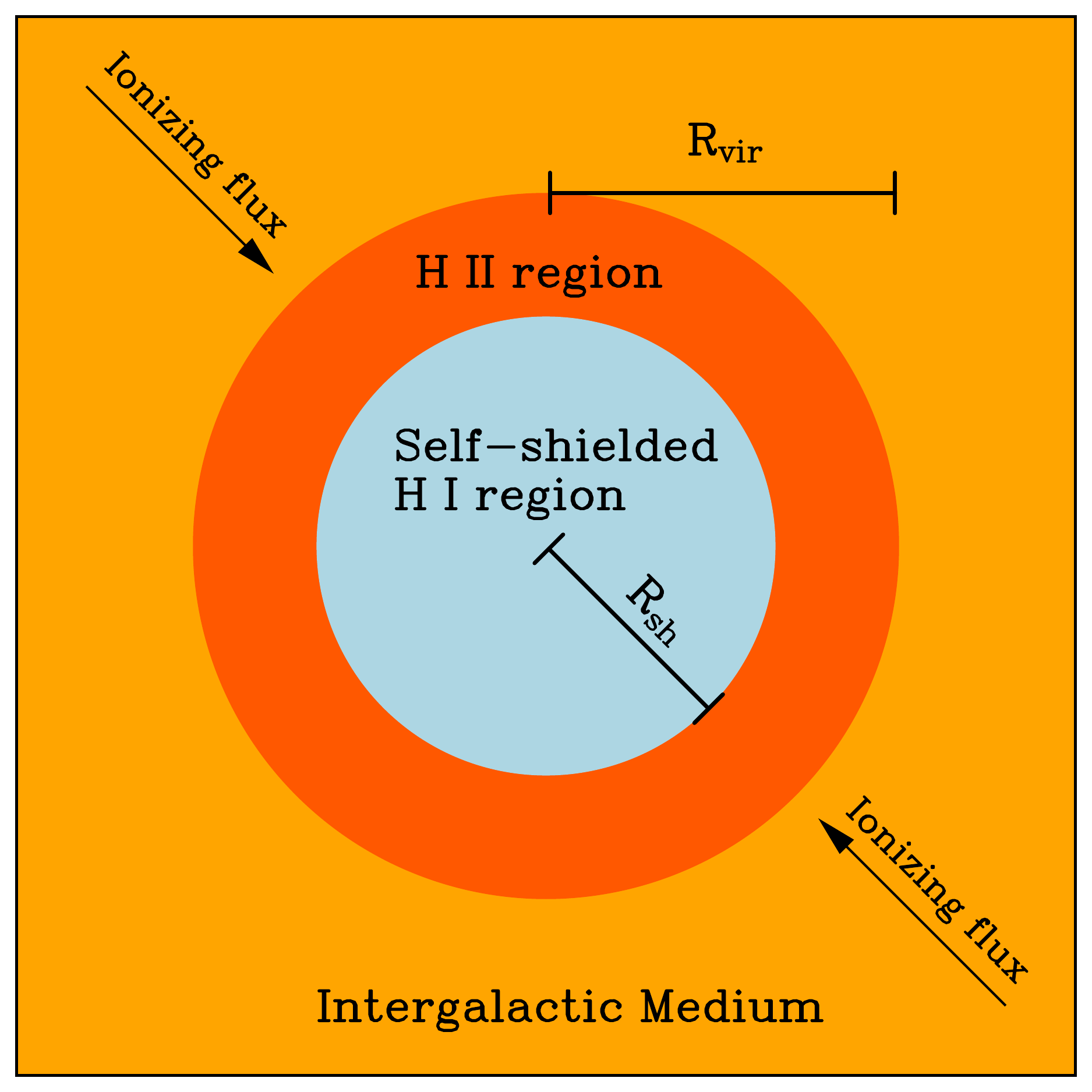}
    \caption{A schematic figure of a halo of radius $\Rvir$ immersed in a uniform photoionizing background. There is potentially a self-shielded H I region of radius $R_{\rm sh}$ within which the gas could cool, collapse, and form stars. }
    \label{fig:ReionSelfShield}
\end{figure}

Several authors have studied the photoevaporation rate of gas clouds subjected to a plane-parallel photoionizing flux \citep[e.g.][]{Bertoldi1989, Mellema1998, Shapiro2004, Iliev2005, Nakatani2020}. For the sake of analytical tractability, and for our interest in a uniform cosmic ionizing background, we instead consider an isotropic photoionizing flux $F_{\rm ion}$ (in $\textrm{cm}^{-2} ~ \textrm{s}^{-1}$). For a given halo immersed in this ionizing background, there is potentially a central self-shielded H I region of radius $R_{\rm sh}$, as sketched in Fig. \ref{fig:ReionSelfShield}. For halo masses below $M_{\rm crit, bound}$, 
the gas is gravitationally unbound in the \HII region of the halo. The resulting rate of gas mass loss from the self-shielded region is \citep[][]{Mellema1998, Nakatani2020}:
\begin{align}
    \Dot{M}_{\rm sh} &= - \mu m_{\rm H}  \int_{\partial V_{\rm sh}} \textrm{d}\boldsymbol{S} \boldsymbol{\cdot} \boldsymbol{\hat{r}} F_{\rm ion} \nonumber \\
    &= - \mu m_{\rm H} \hspace{1 pt} 4 \pi R_{\rm sh}^2 F_{\rm ion}(R_{\rm sh}) \hspace{1 pt} ,
\end{align}
where $F_{\rm ion}(R_{\rm sh})$ is the ionizing flux at radius $r = R_{\rm sh}$. The first line in this expression is practically the same as eq. (15) in \citet{Nakatani2020}, but our assumption of a uniform photoionizing background simplifies the evaluation of the integral considerably. Since $\Dot{M}_{\rm sh} = 4 \pi R_{\rm sh}^2 \rho_{\rm gas}(R_{\rm sh}) \hspace{1 pt} \Dot{R}_{\rm sh}$, this gives us an equation for the radial evolution of the self-shielded region:
\begin{equation}
    \Dot{R}_{\rm sh} = - \frac{F_{\rm ion}(R_{\rm sh})}{n_{\rm gas}(R_{\rm sh})} \hspace{1 pt} \label{Rshdot},
\end{equation}
where $n_{\rm gas} = \rho_{\rm gas}/\mu m_{\rm H}$ is the gas number density in the self-shielded \HI region. If $F_{\rm ion}(R_{\rm sh}) > n_{\rm gas}(R_{\rm sh}) u_{\rm R}$, the ionization front is R-type and therefore 
$F_{\rm ion}(R_{\rm sh})\simeq 2 n_{\rm gas}(R_{\rm sh}) c_{\rm s,HII}$ \citep[e.g.][]{Spitzer1978, Bertoldi1989, Draine2011}. From Eq. (\ref{Rshdot}), this is equivalent to $\vert \Dot{R}_{\rm sh} \vert \gtrsim 2 c_{\rm s,HII}$. When the ionization front approaches the inverse Strömgren layer \citep{Shapiro2004}, the ionizing flux at the front is reduced so much by recombinations along the line of sight that the flux falls below $2 n_{\rm gas}(R_{\rm sh}) c_{\rm s,HII}$ and the front speed slows down to less than $2 c_{\rm s,HII}$. At this point the ionization front transitions from R- to D-type, and travels into the remaining neutral gas of the minihalo (if any) at a slower rate \citep[e.g.][]{Shapiro2004, Nakatani2020}. We investigate the two regimes below.

\subsubsection{Initial R-type ionization front}
\label{RtypeFrontsection}

Here we determine the evolution of the radius $R_{\rm sh}$ of the self-shielded region during the initial R-type phase. The use of the initial gas density profile is justified in this case, because the ionization front during the R-type phase moves at a supersonic velocity $> 2 c_{\rm HII}$, so there is not enough time for the gas to react to its presence. Furthermore, the flux at $r = R_{\rm sh}$ is simply $F_{\rm ion}$ minus the rate of recombinations per unit area in the  column stretching over all $r > R_{\rm sh}$ \citep[e.g.][]{Spitzer1978, Bertoldi1989, BertoldiMcKee1990, Nakatani2020}. Thus, we have \begin{align}
    F_{\rm ion} - F_{\rm ion}(R_{\rm sh}) &= \int_{R_{\rm sh}}^{\infty} \textrm{d}r \hspace{1 pt} k_{1} n_{\rm H^+}^2  \nonumber \\
    &\simeq \int_{R_{\rm sh}}^{\Rvir} \textrm{d}r \hspace{1 pt} \frac{0.857 \hspace{1 pt} k_{1} n_{\rm core}^2 }{[1 + (r/R_{\rm core})^2]^2} \nonumber \\
    &= \frac{0.857 \hspace{1 pt} k_{1} n_{\rm core}^2 \Rvir}{20}  \label{Fioncolumn} \\ &\times \biggl[ \frac{10}{101} + \arctan(10) \nonumber \\ &- \frac{10 (R_{\rm sh}/\Rvir)}{100 ( R_{\rm sh}/\Rvir)^2 + 1} - \arctan\left(10 \frac{R_{\rm sh}}{\Rvir} \right) \biggr] \hspace{1 pt}. \nonumber
\end{align}
The R-type ionization front transitions to D-type near the inverse Strömgren layer radius $R_{\rm ISL}$, defined by $F_{\rm ion}(R_{\rm ISL}) \equiv 0$ \citep{Shapiro2004}. We find that the R-type ionization front can reach $R_{\rm ISL}/\Rvir = 0.566$ (and so photoevaporate the majority of the gas in the halo) for $F_{\rm ion}/k_1 n_{\rm core}^2 \Rvir = 1.2 \times 10^{-4}$. This corresponds to a halo mass $M_{\rm sh, R}$ given by
\begin{equation}
    M_{\rm sh, R} \simeq 5.5 \times 10^9 \left( \frac{F_{\rm ion}}{F_0} \right)^3 \left( \frac{1+z}{10} \right)^{-15} ~ \MSUN \hspace{1 pt} ,
    \label{M_shR}
\end{equation}
where we have normalized the ionizing photon flux to $F_0 = 6.7 \times 10^6 ~ \rm cm^{-2} ~ s^{-1}$ following \citet{Visbal2017}, and have assumed a gas temperature $T_{\rm HII} = 10^4 ~ \rm K$ in the wind region \citep[so that $k_{1} = 2.59 \times 10^{-13} ~ \textrm{cm}^{3} ~ \textrm{s}^{-1}$, see e.g. Table 14.1 in][]{Draine2011}. The above expression has the same scaling found by \citet{Shapiro2004}, but with a different normalization. This is primarily because these authors wanted to calculate the halo mass below which an R-type ionization front photoevaporates \textit{all} the gas in the halo. In contrast, we have calculated the halo mass below which more than half of the gas is photoevaporated -- a more relevant halo mass threshold when considering the star formation efficiency. Furthermore, if the initial R-type ionization front evaporates more than half of the gas in the halo, this will speed up the eventual evaporation by the subsequent D-type ionization front -- the mass scale $M_{\rm sh,D}$ of which we estimate below. 

\subsubsection{Late-time D-type ionization front}
\label{Dtypefrontmass}

If the halo can trap the R-type ionization front, it will turn from R-type to D-type. During this phase the ionization front slows down markedly, producing a shock travelling ahead of the ionization front into the halo \citep[e.g.][]{Ahn2007}. We can model the evolution of $R_{\rm sh}$ during this phase following earlier work on the D-type expansion of H II regions around massive stars \citep[e.g.][]{Spitzer1978, Dyson1980, Raga2012, Bisbas2015, Geen2015}. If the neutral shocked layer is sufficiently thin, pressure equilibrium between this layer and the external H II region can be assumed. Since the pressure in the shocked layer is $P_{\rm ps} = \rho_{\rm HI} \Dot{R}_{\rm sh}^2$, where $ \rho_{\rm HI}$ is the density of gas in the self-shielded region just ahead of the shock, this yields
\begin{equation}
    \Dot{R}_{\rm sh} \simeq - c_{\rm s,HII} \left( \frac{\rho_{\rm HII}}{\rho_{\rm HI}} \right )^{1/2} \hspace{1 pt},
\end{equation}
where we have neglected the gas pressure in the self-shielded region to keep the problem analytically tractable \citep[see e.g.][for the more general case]{Raga2012}. For halo masses below $M_{\rm crit, bound}$, given in Eq. (\ref{reion_noshield}), and during the slow D-type phase, we expect the gas in the H II region to have had enough time to set up a wind profile. If one assumes that the wind velocity is close to the sound speed, then the continuity equation gives $n_{\rm H^+}(r>R_{\rm sh}) = [F_{\rm ion}(R_{\rm sh})/c_{\rm s,HII}](R_{\rm sh}/r)^2$ \citep{Spitzer1978, Bertoldi1989, Nakatani2020}. Since $n_{\rm H^+} \simeq n_{\rm H}$ in the H II region, this yields $\rho_{\rm HII} \simeq \mu_{\rm HII} m_{\rm H} n_{\rm H^+}/0.926$, or $\rho_{\rm HII} \simeq 0.663 n_{\rm H^+} m_{\rm H}$ for a mean molecular weight $\mu_{\rm HII} = 0.614$ in the H II region. If the photoevaporation time-scale is shorter than $\max[t_{\rm cool}, t_{\rm ff}]$ -- as needed for photoionization feedback to be important -- we can assume that the gas density in the H I region is given by the initial density profile, $\rho_{\rm HI}(r < R_{\rm sh}) = \mu_{\rm HI} m_{\rm H} n(r)$, with $\mu_{\rm HI} = 1.23$ and $n(r)$ from Eq. (\ref{nprofile}). Taken together, this yields 
\begin{equation}
    \Dot{R}_{\rm sh} \simeq - 0.734 \hspace{1 pt} c_{\rm s,HII} \left[ \frac{F_{\rm ion}(R_{\rm sh})}{c_{\rm s,HII} n_{\rm core}} \right ]^{1/2} \left[1 + 100 \left( \frac{R_{\rm sh}}{\Rvir}\right)^2 \right]^{1/2} \hspace{1 pt} \label{Rshdot_Dtype}.
\end{equation}
Following Eq. (\ref{Fioncolumn}), the ionizing flux at $R_{\rm sh}$ in this case, with the assumed wind profile, is determined by \citep[][]{Spitzer1978, Bertoldi1989}:
\begin{align}
    F_{\rm ion} - F_{\rm ion}(R_{\rm sh}) &= \int_{R_{\rm sh}}^{\infty} \textrm{d}r \hspace{1 pt} k_{1} n_{\rm H^+}^2  \nonumber \\
    &\simeq k_1 \left[ \frac{F_{\rm ion}(R_{\rm sh})}{c_{\rm s,HII}} \right]^2 \int_{R_{\rm sh}}^{\infty} \textrm{d}r \hspace{1 pt} \left( \frac{R_{\rm sh}}{r} \right)^4 \label{FionDtype} \\
    &=  \frac{k_1}{3} \left[ \frac{F_{\rm ion}(R_{\rm sh})}{c_{\rm s,HII}} \right]^2 R_{\rm sh} \hspace{1 pt}. \nonumber
\end{align}
We solve the quadratic equation for $F_{\rm ion}(R_{\rm sh})$ and insert the result into Eq. (\ref{Rshdot_Dtype}) to get 
\begin{align}
    \Dot{R}_{\rm sh} &\simeq - 0.734 \hspace{1 pt} c_{\rm s,HII} \left(\frac{3 c_{\rm s,HII}}{2 k_1 n_{\rm core} R_{\rm sh}} \right)^{1/2} \left[1 + 100 \left( \frac{R_{\rm sh}}{\Rvir} \right)^2 \right]^{1/2} \nonumber \\ &\times \left( \sqrt{1 + \frac{4 k_1 F_{\rm ion}}{3 c_{\rm s,HII}} R_{\rm sh}} - 1 \right)^{1/2} \hspace{1 pt} \label{Rshdot_DtypeII}.
\end{align}
At this point it is convenient to introduce dimensionless variables:
\begin{equation}
    \xi \equiv \frac{R_{\rm sh}}{\Rvir} \hspace{1 pt}, \hspace{3 pt} \Tilde{t} \equiv \frac{ 0.734 \hspace{1 pt} c_{\rm s,HII} t}{\Rvir} \left( \frac{3 c_{\rm s,HII}}{2 k_1 n_{\rm core} \Rvir} \right )^{1/2} \hspace{1 pt}, \label{dimensionlessevap}
\end{equation}
and 
\begin{equation}
    q \equiv \frac{4 k_1 F_{\rm ion} \Rvir}{3 c_{\rm s,HII}^2} \hspace{1 pt} .
\end{equation}
This variable $q$ is a key parameter of the model and, except for the trivial numerical factor $4/3$, also appears in \citet{Bertoldi1989} (as $\psi$) and \citet{Nakatani2020} (also as $q$). With the above dimensionless variables, Eq. (\ref{Rshdot_DtypeII}) simplifies to
\begin{equation}
    \frac{\textrm{d}\xi}{\textrm{d}\Tilde{t}} \simeq - \frac{(1 + 100 \xi^2)}{\xi^{1/2}} \left( \sqrt{1 + q\xi} - 1 \right)^{1/2} \hspace{1 pt} .
\end{equation}
Thus, the (dimensionless) photoevaporation time $\Tilde{t}_{\rm evap}$ and half-mass photoevaporation time $\Tilde{t}_{\rm evap,1/2}$ can be found by integration:
\begin{align}
    \Tilde{t}_{\rm evap} &\simeq \int_{0}^{1}  \frac{\textrm{d}\xi \sim \xi^{1/2}}{(1+100 \xi^2)(\sqrt{1 + q\xi} - 1)^{1/2}} \hspace{1 pt} \label{tevap},\\
    \Tilde{t}_{\rm evap,1/2} &\simeq \int_{\xi_{1/2}}^{1} \frac{\textrm{d}\xi \hspace{1 pt} \xi^{1/2}}{(1+100 \xi^2)(\sqrt{1 + q\xi} - 1)^{1/2}} \hspace{1 pt} \label{tevap_half}.
\end{align}
Here $\xi_{1/2}$ is the self-shielding radius containing half of the initial gas mass. 

In the simplified case of a uniform photoionizing background (see Fig. \ref{fig:ReionSelfShield}), the boundary of the self-shielded region is spherically symmetric, and for the assumed initial density profile we get $\xi_{1/2} = 0.566$ (see Sec. \ref{sec:Halo properties}). In the other case of an assumed plane-parallel ionizing source, the effective $\xi_{1/2}$ is expected to be slightly smaller because the side facing the ionizing source casts a shadow \citep[see e.g.][]{Nakatani2020} that encloses an approximately cylindrical neutral gas mass larger than expected for a spherical region of radius $\xi$. Given the approximations made so far, we simply adopt $\xi_{1/2} = 0.5$ as a rough estimate. The two integrals in Eqs. (\ref{tevap_half}) and (\ref{tevap_half}) are then approximately given by:
\begin{align}
    \Tilde{t}_{\rm evap}(q) ~ &\simeq \left[ \left(\frac{0.42401}{q^{1/2}}\right)^b + \left(\frac{0.201913}{q^{1/4}}\right)^b \right ]^{1/b} \hspace{1 pt} \label{tevap_approx1},\\
    \Tilde{t}_{\rm evap,1/2}(q) ~ &\simeq \left[ \left(\frac{0.096985}{q^{1/2}}\right)^b + \left(\frac{0.062978}{q^{1/4}}\right)^b \right ]^{1/b} \hspace{1 pt} \label{tevap_half_approx1},
\end{align}
where $b = 2.7$. Both analytical approximations to the integrals have the correct asymptotic limits for $q \rightarrow 0$ and $q \rightarrow \infty$, and the expression for $\Tilde{t}_{\rm evap}$ ($\Tilde{t}_{\rm evap,1/2}$) is accurate to within $1.7\%$ ($2.9\%$) for all $q$. 
\begin{figure}
    \includegraphics[trim={0.1cm 0.4cm 0.4cm 0.4cm},clip,width = \columnwidth]{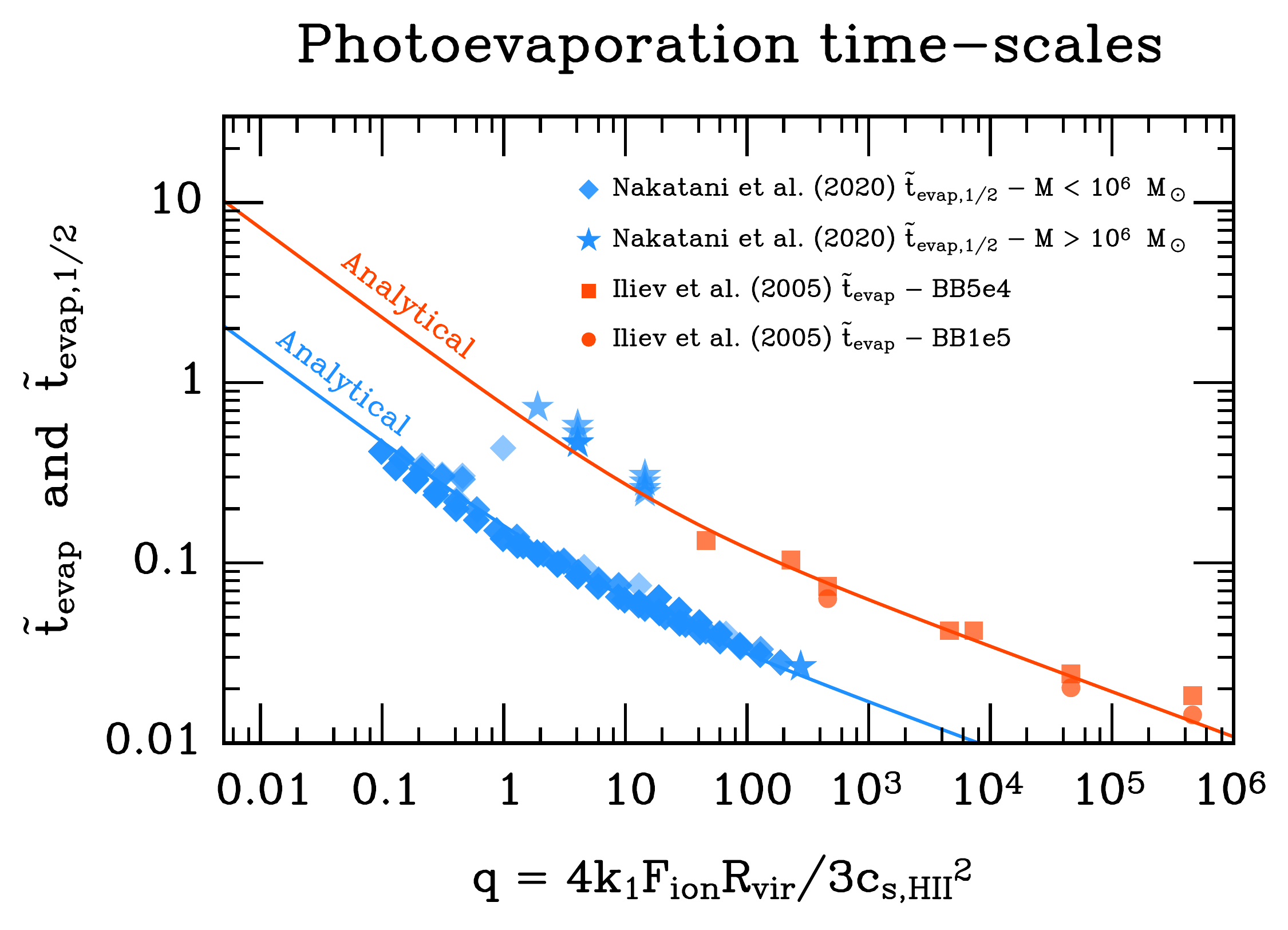} \caption{A comparison between the predicted dimensionless evaporation time-scales $\Tilde{t}_{\rm evap,1/2}$ (bottom blue line) and $\Tilde{t}_{\rm evap}$ (top orange line), multiplied by fudge factors of 1.5 and 1.7, respectively, and simulation data from \citet{Iliev2005} (orange scatter points) and \citet{Nakatani2020} (blue scatter points).$^{\ref{tevapfootnote}}$ Some of the minihaloes with masses $\ge 10^6 ~ \MSUN$ in the simulations of \citet{Nakatani2020} (star symbols) undergo H$_2$-cooling and condense before significant photoevaporation can take place, changing the internal density structure and boosting the photoevaporation time-scale relative to the model prediction that assumed the initial pre-collapse density profile.
    }
    \label{fig:tevap}
\end{figure}

The modelling makes use of several simplifying approximations --- e.g. we neglected the gas pressure in the H I region, assumed a constant wind velocity $c_{\rm s,HII}$ --- so we expect that Eqs. (\ref{tevap_approx1}) and (\ref{tevap_half_approx1}) are accurate to within a factor of a few. However, the key result from the model is that the evaporation time scales, when normalized as in Eq. (\ref{dimensionlessevap}), should only be a function of $q$. In particular, scaling as $q^{-1/2}$ for small $q$, and $q^{-1/4}$ for large $q$. In Fig. \ref{fig:tevap}, we compare this analytical prediction with numerical simulations of photoevaporating minihaloes exposed to a plane-parallel ionizing background \citep{Iliev2005, Nakatani2020}.\footnote{The data from \citet{Nakatani2020} include all their haloes that have at most $1\%$ of their initial gas mass remaining (i.e. $C_{1} < 0.01$ in their eq. 28). The data from \citet{Iliev2005} is taken from their highest resolution simulations ($1024 \times 2048$) assuming $5 \times 10^4 ~ \rm K$ (BB5e4) and $10^5 ~ \rm K$ (BB1e5) blackbody spectra, as listed in their tables 1 and 2. The data points from \citet{Nakatani2020} plotted in Fig. \ref{fig:tevap} cover halo masses $10^3 ~ \MSUN \leq M \leq 3.16 \times 10^6 ~ \MSUN$, redshifts $10 \leq z \leq 20$, ionizing fluxes $8.1 \times 10^3  ~ \textrm{cm}^{-2} ~ \textrm{s}^{-1} \leq F_{\rm ion} \leq 8.1 \times 10^5 ~ \textrm{cm}^{-2} ~ \textrm{s}^{-1}$, and gas metallicities $0 \leq Z/Z_{\odot} \leq 10^{-3}$ with H$_2$-cooling included. The data points from \citet{Iliev2005} cover a halo mass range $10^3 ~ \MSUN \leq M \leq 4 \times 10^7 ~ \MSUN$ (although all except two points have $M=10^7 ~ \MSUN$), ionizing fluxes $4.18 \times 10^5 ~ \textrm{cm}^{-2} ~ \textrm{s}^{-1} \leq F_{\rm ion} \leq 8.36 \times 10^8 ~ \textrm{cm}^{-2} ~ \textrm{s}^{-1}$, a single redshift $z = 9$, and gas metallicity $Z/Z_{\odot} = 10^{-3}$ with no H$_2$-cooling. \label{tevapfootnote}} 

We see that there indeed is a tight correlation between the dimensionless evaporation time scales and $q$, with the slope being consistent with our analytical prediction over 6 orders of magnitude in $q$. The time-scales in Eq. (\ref{tevap}) and (\ref{tevap_half}) are off by less than a factor of $2$ when compared to the simulations, which can be fixed by a fudge factor of $1.5$ for $\Tilde{t}_{\rm evap,1/2}$, and $1.7$ for $\Tilde{t}_{\rm evap}$, as was used in Fig. \ref{fig:tevap}. We note that both \citet{Iliev2005} and \citet{Nakatani2020} provided fits to the evaporation time-scales from their simulations, and \citet{Nakatani2020} even suggested a dependence on $q$ (but did not predict the broken power law analytically). These fits have very similar dependencies on $M$ and $F_{\rm ion}$ as our analytical expressions, but do have a slightly stronger redshift dependence. However, because of the relatively small redshift range probed by these authors, the plausible physical basis for the model in this section, and the very good agreement with their simulation data, we adopt the model presented here. 

Thus, to proceed we take the half-mass photoevaporation time-scale to be given by Eq. (\ref{tevap_half_approx1}) with a fudge factor $\simeq1.5$:
\begin{align}
    t_{\rm evap, 1/2} ~ &\simeq \frac{2 \Rvir}{c_{\rm s, HII}} \left( \frac{2 k_1 n_{\rm core} \Rvir}{3 c_{\rm s,HII}} \right)^{1/2} \nonumber \\ &\times \left[ \left(\frac{0.096985}{q^{1/2}}\right)^b + \left(\frac{0.062978}{q^{1/4}}\right)^b \right ]^{1/b} \hspace{1 pt} \label{tevap_half_final}. 
\end{align}
 As noted and found by \citet{Nakatani2020}, gas cooling and collapse can take place in minihaloes if photoevaporation is sufficiently slow. Gas can condense on a time-scale $\max[t_{\rm cool}, t_{\rm ff}]$, so we only expect D-type photoevaporation to be important when $t_{\rm evap, 1/2} < \max[t_{\rm cool}, t_{\rm ff}]$. A necessary condition for star formation in a photoionizing background is, therefore, $t_{\rm evap, 1/2} > t_{\rm ff}$. This will only be satisfied for haloes above a certain mass $M_{\rm sh,D}$ that have long enough photoevaporation times. In Appendix \ref{photoevaporationscales} we show that if we set Eq. (\ref{tevap_half_final}) equal to $\tff$ we can solve for the corresponding halo mass $M_{\rm sh,D}$. The final result is 
\begin{align}
    M_{\rm sh,D} ~ &\simeq ~ \frac{4 \pi}{3} \frac{387 (F_{\rm ion} c_{\rm s,HII}/n_{\rm core})^{3/2} t_{\rm ff}^3 \rhovir}{ \left[1 + 1.15^\gamma \left( \dfrac{k_1^2 F_{\rm ion}^3 t_{\rm ff}^2}{n_{\rm core} c_{\rm s,HII}^3} \right)^{\gamma/8}  \right]^{12/5\gamma}} \nonumber \\
    ~ &= ~  \frac{1.37 \times 10^8 ~ \MSUN}{ \left[1 + 120 \left ( \dfrac{F_{\rm ion}}{F_0} \right )^{0.975} \left ( \dfrac{1+z}{10} \right )^{-1.95}  \right]^{0.923}} \label{M_shD}\\
    ~ &\times ~ \left ( \frac{F_{\rm ion}}{F_0} \right )^{3/2} \left ( \frac{1+z}{10} \right )^{-6}  \hspace{1 pt} \nonumber,
\end{align}
where $\gamma = 2.6$. On the second line we have, as in Sec. \ref{RtypeFrontsection}, assumed a gas temperature $T_{\rm HII} = 10^4 ~ \rm K$ in the wind region so that $c_{\rm s,HII} = 1.16 \times 10^6 ~ \textrm{cm s}^{-1}$ and $k_{1} = 2.59 \times 10^{-13} ~ \textrm{cm}^{3} ~ \textrm{s}^{-1}$ \citep{Draine2011}, $t_{\rm ff}$ from Eq. (\ref{tff}), and $n_{\rm core}$ from the high-mass limit in Eq. (\ref{ncore}). We have also normalized the ionizing photon flux to $F_0 = 6.7 \times 10^6 ~ \rm cm^{-2} ~ s^{-1}$ as in Eq. (\ref{M_shR}).

\subsubsection{Summary of photoionization feedback}
\label{reionsummary}

To summarize all the effects of photoionization feedback, we take the critical halo mass $M_{\rm crit, reion}$ above which the gas in most haloes can evade photoevaporation in a photoionizing background $F_{\rm ion}$ to be given by
\begin{equation}
    M_{\rm crit, reion}(z, F_{\rm ion}) = \textrm{min}\left[ \textrm{max}\left(M_{\rm sh,R}, M_{\rm sh,D} \right), \hspace{1 pt} M_{\rm crit,bound} \right] \hspace{1 pt},
    \label{reionsummarycritmass}
\end{equation}
with $M_{\rm crit,bound}$ given by Eq. (\ref{reion_noshield}), $M_{\rm sh,R}$ by Eq. (\ref{M_shR}) and $M_{\rm sh,D}$ by Eq. (\ref{M_shD}).

\subsection{Baryon streaming velocities}
\label{sec:streaming}

\begin{figure*}
\includegraphics[width=0.8\textwidth]{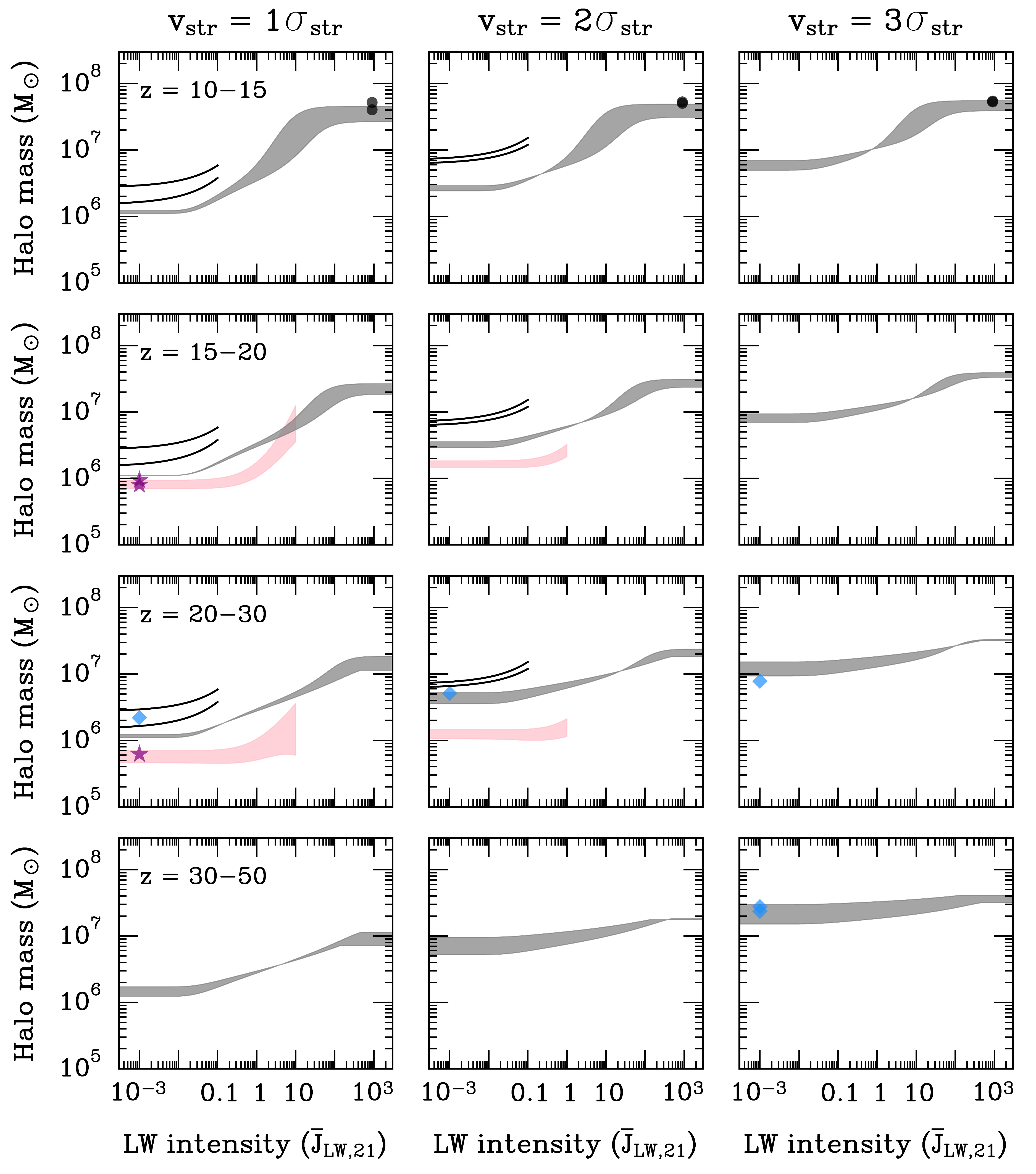}
\caption{The cooling threshold as a function of the Lyman-Werner (LW) intensity $\J$ for different streaming velocities $v_{\rm str}$ (columns) and redshift $z$ ranges (rows), assuming no photoionization feedback. The grey bands are the predicted thresholds for a given redshift range from our model. Individual minihaloes from cosmological simulations are shown with symbols: purple stars are from \citet{Greif2011}, black circles from \citet{Latif2014_streaming}, and blue diamonds from \citet{Hirano2018}. If the minihalo comes from a  simulation with $\J = 0$ we have placed its symbol at $\J = 10^{-3}$ instead. The black lines show the redshift-independent thresholds ($M_{\rm min}$ and $M_{\rm ave}$) which \citet{Schauer2021} fitted to their simulation results, for the regimes where these fits are valid (they break down for $v_{\rm str} = 3 \sigma_{\rm str}$ because of a small sample). The pink bands are the redshift-dependent fits to the cooling threshold $M_{50\%}$ from \citet{Kulkarni2021}, also only plotted in the regimes where their fits are valid.
}
\label{BSV}
\end{figure*}

So far we have only considered cooling processes and the effect of radiative feedback. However, there is an additional effect that can suppress gas condensation in low-mass haloes, namely, baryon streaming velocities (BSV) \citep[e.g.][]{Tseliakhovich2010, Fialkov2012}. Until the epoch of recombination ($z_{\rm rec} \simeq 1000$), baryons were tightly coupled to the photons. The resulting radiation pressure stabilized the baryons against self-collapse, while the DM could start to clump together. When the baryons decoupled from the photons, the result was a relative velocity difference $v_{\rm str}$ between the baryons and DM, which we refer to as the BSV. The BSV are coherent on comoving scales of $\lesssim \textrm{few} \times \textrm{Mpc}$, and have a Maxwell-Boltzmann distribution. The RMS streaming velocity $\sigma_{\rm str}$ at redshifts $z < z_{\rm rec}$ is given by \citep[e.g.][]{Tseliakhovich2010, Fialkov2012}
\begin{equation}
\sigma_{\rm str}(z) \simeq 30 \hspace{1 pt} \left( \frac{1+z}{1000} \right) ~ \textrm{km s}^{-1} \hspace{1 pt}.
\end{equation}
Since all streaming velocities decay as $v_{\rm str} \propto (1+z)$, the BSV at a given position will be a constant multiple of $\sigma_{\rm str}(z)$. 

At high redshifts, the baryons could move sufficiently rapidly to escape the potentials of low-mass dark matter haloes, hence suppressing gas accretion onto these haloes \citep[e.g.][]{Stacy2011,Greif2011,Schauer2019,Schauer2021,Kulkarni2021}. Based on simple physical arguments and the results of simulations, \citet{Fialkov2012} proposed that BSV modify the cooling threshold according to:
\begin{align}
    v_{\rm crit} = \sqrt{v_{\rm crit, nostr}^2 + [\alpha v_{\rm str}]^2} \hspace{1 pt} \label{Fialkovmodel},
\end{align}
where $v_{\rm crit}$ and $v_{\rm crit, nostr}$ are the critical virial velocities above which most haloes form dense gas, with and without streaming velocities, respectively. The effect of the streaming velocities is captured by the term $\alpha v_{\rm str}$, where $\alpha > 1$ is a constant that encapsulates the fact that the streaming velocities are larger during the assembly of the halo. For very large BSV, Eq. (\ref{Fialkovmodel}) implies that gas can only accrete onto DM haloes with virial velocities $\vvir \gtrsim \alpha v_{\rm str}$, roughly consistent with a simple Jeans scale argument \citep[][]{Stacy2011}. In the limit of low BSV, gas condensation only takes place in haloes with $\vvir \gtrsim v_{\rm crit, nostr}$, with $v_{\rm crit, nostr}$ determined by the gas cooling rate, as discussed in the previous sections.

On the basis of the results of the cosmological simulations of \citet{Stacy2011} and \citet{Greif2011}, \citet{Fialkov2012} adopted a redshift independent value $v_{\rm crit, nostr} = 3.714 ~ \mathrm{km s}^{-1}$, along with $\alpha = 4.015$. Only a handful of haloes were studied by \citet{Stacy2011} and \citet{Greif2011}, and on top of this, \citet{Greif2011} argue that the lower resolution used in the simulations of \citet{Stacy2011} leads to an overestimate of the cooling threshold. Because of this, and the much greater sample of haloes available at the present-day for comparisons, we update values of $v_{\rm crit, nostr}$ and $\alpha$ in the model of \citet{Fialkov2012}. In the absence of BSV, we have already calculated the cooling threshold in the previous sections and shown it to be in good agreement with a number of high-resolution cosmological simulations. Thus, we take (using Eq. \ref{vvir}):
\begin{equation}
    v_{\rm crit, nostr} = 3.65 \hspace{1 pt} \left( \frac{M_{\rm crit, nostr}}{10^6 ~ \MSUN} \right)^{1/3} \Z^{1/2} ~ \textrm{km s}^{-1} \hspace{1 pt} ,\label{nostreamvcrit}
\end{equation}
with $M_{\rm crit,nostr} = \textrm{max}(M_{\rm crit, H_2+LW}, \hspace{1 pt} M_{\rm crit, reion})$ being the cooling threshold in the absence of streaming velocities. The value of $\alpha$ is set to 6 in order to obtain a better fit for the simulation data.

Fig. \ref{BSV} shows the predicted cooling thresholds of Eqs. (\ref{Fialkovmodel}) and (\ref{nostreamvcrit}) and a comparison to high-resolution cosmological simulations \citep[][]{Greif2011, Latif2014_streaming, Hirano2018_BSV, Kulkarni2021, Schauer2021}. Since we are comparing to simulations without a photoionizing background, we have set $F_{\rm ion} = 0$, so that $M_{\rm crit,nostr} = M_{\rm crit, H_2+LW}$. As before in Figs. \ref{H2coolingfigure} \textbf{and \ref{fig:J21}}, there is a disagreement between \citet{Schauer2021} and \citet{Kulkarni2021}, with our model predictions and the other handful of simulations falling somewhere between the two for low $\J$. The lack of many independent high-resolution simulations with baryon streaming velocities, unfortunately, prevents a more detailed comparison. A more detailed model and comparison have to await further simulation data. We note though that at lower redshifts the effect of BSV is expected to become less important in determining $M_{\rm crit}$ since BSV decay as $\propto (1+z)$, and because of the increasing importance of LW and photoionizing feedback. This was recently borne out in the simulations of \citet{Schauer2022Fire3}, who used FIRE-3 simulations of galaxy formation with BSV added. They found that $v_{\rm str} = 2\sigma_{\rm str}$ increases $M_{50\%}$ by a factor $\sim 2$ at $z = 10$, but has a negligible impact by $z = 5$.

\subsection{Summary of cooling model}
\label{sec:coolingsummary}

To summarize all of the previous calculations, efficient gas cooling and collapse is expected to take place in the majority of haloes of mass $M > M_{\rm crit}$, where:
\begin{align}
    M_{\rm crit}(z, \J, F_{\rm ion}, v_{\rm str}) &= ~ 2.06 \times 10^{4} ~ \MSUN \\ &\times \hspace{1 pt} \left( v_{\rm crit, nostr}^2 + [\alpha v_{\rm str}]^2 \right )^{3/2} \nonumber \hspace{1 pt} ,
    \label{Mcrit_final}
\end{align}
where $\alpha = 6$ is a constant. The virial velocity $v_{\rm crit, nostr}$ corresponding to the critical halo mass $M_{\rm crit,nostr}$ for cooling in the absence of streaming velocities is given by
\begin{align}
    v_{\rm crit, nostr}(z, \J, F_{\rm ion}) &= 3.65 \hspace{1 pt} \left( \frac{M_{\rm crit,nostr}}{10^6 ~ \MSUN} \right)^{1/3} \nonumber \\ &\times \Z^{1/2} ~ \textrm{km s}^{-1} \hspace{1 pt} , \\
    M_{\rm crit,nostr}(z, \J, F_{\rm ion}) &= \textrm{max}\left[M_{\rm crit, H_2+LW}, \hspace{1 pt} M_{\rm crit, reion}\right] \hspace{1 pt},
\end{align}
with $M_{\rm crit, H_2+LW}$ and $M_{\rm crit,reion}$ given by Eqs. (\ref{M_H2LW}) and (\ref{reionsummarycritmass}), respectively. 

\begin{figure*}
\includegraphics[trim={0.2cm 0.5cm 0.4cm 0.4cm},clip,width=0.8\textwidth]{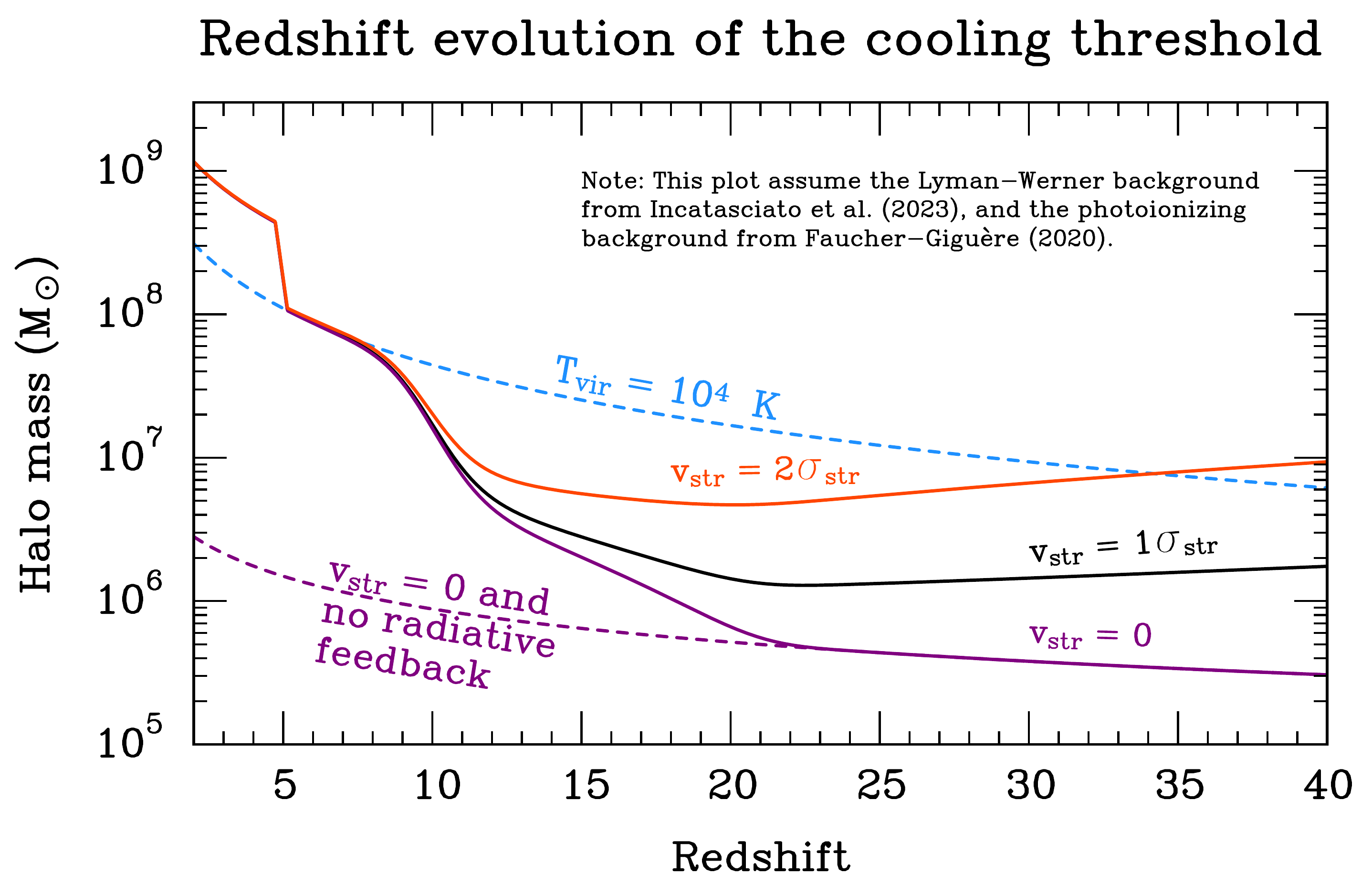}
\caption{A plot of the predicted redshift evolution of the cooling thresholds (solid lines), assuming the estimated cosmic mean Lyman-Werner (LW) background from \citet{Incatasciato2023_LW}, the photoionizing background from \citet{Faucher2020_UV}, and different baryon streaming velocities (from zero at the bottom to $v_{\rm str} = 2\sigma_{\rm str}$ at the top). The dashed lines indicate the cooling thresholds in the absence of radiative feedback and streaming velocities (bottom dashed line), and the atomic-cooling threshold ($\Tvir = 10^4 ~ \rm K$, top dashed line). The actual cooling threshold is seen to fall somewhere between these two for redshifts $33 \lesssim z \lesssim 5$. Following reionization, the increasing ionizing background and lower gas density in haloes make reionization feedback effective, leading to a sharp uptick in the cooling threshold \citep[compare with the schematic plot in fig. 1 from][]{Benitez2020}. 
}
\label{SummaryCooling}
\end{figure*}

Given the multitude of factors that go into calculating $M_{\rm crit}$, we show the result of an example calculation in Fig. \ref{SummaryCooling}. For this calculation, we have assumed the recently estimated cosmological LW background of \citet{Incatasciato2023_LW} and the photoionizing background from \citet{Faucher2020_UV}.\footnote{Instead of numerically integrating the spectrum from \citet{Faucher2020_UV} at each redshift to get $F_{\rm ion}$, we estimate $F_{\rm ion}$ as follows. The photoionization rate is $\Gamma_{\rm ion} = \int_{\nu_{\rm LyC}}^{\infty} \textrm{d}\nu \hspace{1 pt} 4 \pi J_{\nu} \sigma_{\rm ion}/h\nu$, where $h \nu_{\rm LyC} = 13.6 ~ \rm eV$ and $\sigma_{\rm ion}$ is the photoionization cross-section. Written in terms of the flux-averaged photoionization cross-section $\Bar{\sigma}_{\rm ion}$ this becomes $\Gamma_{\rm ion} = \Bar{\sigma}_{\rm ion} \int_{\nu_{\rm LyC}}^{\infty} \textrm{d}\nu \hspace{1 pt} 4 \pi J_{\nu}/h\nu$. Since the photoionizing photon flux on the "surface" of a spherical halo from a uniform background is $F_{\rm ion} = \int_{\nu_{\rm LyC}}^{\infty} \textrm{d}\nu \hspace{1 pt} 2 \pi J_{\nu}/h\nu$ \citep[e.g.][]{Shapiro2004}, we find $F_{\rm ion} = \Gamma_{\rm ion}/2 \Bar{\sigma}_{\rm ion}$. The ionizing spectrum shortward the Lyman limit from \citet{Faucher2020_UV} can be crudely approximated as a power law $J_{\nu} \propto \nu^{-\alpha}$ with $\alpha \simeq 1$ \citep[see also e.g.][]{Nakatani2020}, from which we get $\Bar{\sigma}_{\rm ion} \simeq 1.5 \times 10^{-18} ~ \rm cm^2$. With this estimate of $\Bar{\sigma}_{\rm ion}$ and the values of $\Gamma_{\rm ion}$ provided by \citet{Faucher2020_UV} we can get an estimate of $F_{\rm ion}$. We note that our final result in Fig. \ref{SummaryCooling} is not sensitive to our choice of $\Bar{\sigma}_{\rm ion}$ -- increasing $\Bar{\sigma}_{\rm ion}$ by a factor of $2$ \citep[appropriate for local galaxies, e.g.][]{Agertz2020_UFD} only move the jump in $M_{\rm crit}$ from $z \simeq 5$ to $z \simeq 4.5$.} The former was calculated using post-processed radiative transfer calculations for the FiBY cosmological simulations that include both Pop III and Pop II star formation. The latter was designed to reproduce, among other things, the observed luminosity functions of galaxies and AGN, and a reionization history consistent with existing observational constraints. For these backgrounds, hydrogen is reionized by $z \simeq 7$, and the LW background grows exponentially from $\J \simeq 0.1$ at $z \simeq 18$ to $\J \simeq 15$ at $z \simeq 6$. 

Assuming the above-mentioned radiation backgrounds, we see the following evolution in Fig. \ref{SummaryCooling}:
\begin{enumerate}
\item $40 > z \gtrsim 20$: For these high redshifts radiative feedback has a small effect on $M_{\rm crit}$, with the cooling threshold being mainly determined by baryon streaming and the efficiency of H$_2$-cooling. In the absence of baryon streaming velocities, $M_{\rm crit} \sim \textrm{few} \times 10^5 ~ \MSUN$, which increases to nearly $10^7  ~ \MSUN$ for rare regions with high streaming velocities ($v_{\rm str} =  2\sigma_{\rm str}$). For a more typical region with $v_{\rm str} =  \sigma_{\rm str}$ we find $M_{\rm crit} \sim 1-2 \times 10^6 ~ \MSUN$ in this redshift range.

\item $20 > z \gtrsim 8$: In this redshift range, the steadily increasing LW background starts to have an increasingly important effect on $M_{\rm crit}$. By $z \sim 10$, the cooling threshold converges to $M_{\rm crit} \sim 1-2 \times 10^7 ~ \MSUN$ regardless of the streaming velocity assumed, very similar to the findings of \citet{Schauer2022Fire3}, who ran FIRE-3 simulations with and without streaming velocities. By $z \sim 8$, we find that the LW feedback is effective enough to suppress cooling in all minihaloes ($\Tvir < 10^4 ~ \rm K$). These results imply that minihalos can contribute substantially to cosmic reionization, roughly up to its midpoint.

\item $8 > z > 2$: By $z \sim 8$, cooling is suppressed in all minihaloes, and reionization is well underway. At first, all haloes above the atomic-cooling threshold ($\Tvir = 10^4 ~ \rm K$) can form stars. Eventually, around $z \sim 5$, the photoionizing background becomes strong enough, and the gas in virialized haloes is not dense enough to self-shield. This lead to an additional sharp jump in the cooling threshold by a factor of $\sim 4$. The cooling threshold does not increase further because photoheated gas can remain gravitationally bound above this limit. Note that the jump in $M_{\rm crit}$ due to reionization feedback is delayed relative to the reionization of the IGM itself because of effective self-shielding.
\end{enumerate}

We want to emphasize that the above summary of the evolution of $M_{\rm crit}$ hinges on the assumed radiation backgrounds. While we regard them as plausible, other backgrounds could change these predictions. Furthermore, in reality, there would be spatial fluctuations in the radiation backgrounds (and the streaming velocities on larger scales). For example, a halo close to a massive galaxy could experience stronger radiative feedback, and hence a boosted cooling threshold at its position. Thus, if one were to implement our model for the cooling threshold in a simulation it would in general be spatially dependent.

\section{Summary and conclusion}
\label{SummaryAndConclusion}

The first stars and galaxies formed in the lowest mass DM haloes in which gas could cool efficiently. The predicted abundance and stellar masses of these luminous objects is a sensitive function of the halo mass cooling threshold $M_{\rm crit}$ above which most haloes could form stars. In this paper, we have calculated $M_{\rm crit}$ in detail analytically and 
semi-analytically taking a wide range of processes into account. We also compared our calculations with numerous available high-resolution cosmological hydrodynamical simulations. Our findings can be summarized as follows:
\begin{itemize}
\item \textit{H$_2$-cooling in minihaloes:} In Sec.~\ref{sec:H2threshold} we analytically derived the critical halo mass above which H$_2$-cooling becomes effective, in the absence of radiative feedback and baryon streaming velocities. In this case, we find that most haloes can host cool gas above $\sim \textrm{few} \times 10^5 - 10^6 ~ \MSUN$, depending on redshift. The derived redshift dependence is significantly weaker than found analytically by \citet{Trenti2009}. This finding is mainly due to our physically more accurate modelling of the density structure in minihaloes. We compare our calculation to the results of many state-of-the-art high-resolution cosmological hydrodynamical simulations and find that it is in better agreement with simulation data than earlier analytical treatments.

\item \textit{Effect of radiative feedback on H$_2$-cooling in minihaloes:} Next we calculated the effect of radiative feedback on the H$_2$-cooling threshold semi-analytically. More specifically, we considered the photodissociation of H$_2$ by LW photons, and the photodetachment of H$^{-}$ by IR photons. Unlike many earlier treatments, we incorporate self-shielding against the LW background and find results consistent with recent high-resolution simulations which include self-shielding. 
In contrast, earlier models that did not incorporate self-shielding dramatically overestimate the effects of LW feedback, as also has been pointed out in recent numerical studies \citep[e.g.][]{Skinner2020, Kulkarni2021, Schauer2021}.

\item \textit{Effect of metals on cooling in minihaloes:} For completeness we also considered the potential effect of metals on the cooling threshold. Metals could boost the cooling rate both directly via metal fine-structure line cooling, and indirectly via dust-catalyzed H$_2$ formation. We find that at the low metallicities of interest, both effects have a negligible effect on the cooling threshold, and are therefore neglected in our final expression for $M_{\rm crit}$. 

\item \textit{Photoionizing feedback:} Next we considered the effect of photoionization feedback, which can photoevaporate the gas in low-mass haloes. We take self-shielding against the ionizing background into account by analytically modelling the evolution of ionization fronts (both R and D-type) in minihaloes, and find good agreement with simulations of photoevaporating minihaloes. Because of self-shielding, which is particularly effective at high redshifts, reionization feedback is expected to be slightly delayed relative to the reionization of the IGM. In particular, $M_{\rm crit}$ by a factor of $\sim 4$ over the atomic-cooling threshold ($\Tvir = 10^4 ~ \rm K$) only by $z \sim 5$. This delay is consistent with cosmological simulations of dwarf galaxies that explicitly include self-shielding \citep[e.g.][]{Jeon2017_UFD, Wheeler2019}, and the observed star formation history of Ultra-Faint Dwarf galaxies \citep{Brown2014, Collins_2022}.

\item \textit{Baryon streaming velocities:} Finally we consider the effect of baryon streaming velocities (BSV) on $M_{\rm crit}$. The BSV mostly affect $M_{\rm crit}$ at high redshift ($z \gtrsim 10-15$, see Fig. \ref{SummaryCooling}) and low LW intensities, consistent with earlier studies. At lower redshifts, BSV has a negligible effect on $M_{\rm crit}$ as a result of decaying streaming velocities ($v_{\rm str} \propto 1+z$) and the increasingly dominant effect of radiative feedback.
\end{itemize}

The final expression for $M_{\rm crit}$, taking molecular and atomic cooling, radiative feedback, and baryon streaming velocities into account, was provided in Eq. (\ref{Mcrit_final}). It can be implemented in any semi-analytical model which needs to establish which haloes can host cool gas, and potentially form stars. The inputs for the calculation of $M_{\rm crit}$ are the redshift, the LW intensity $\J$, the ionizing photon flux $F_{\rm ion}$, and the streaming velocity $v_{\rm str}$. In a semi-analytical framework the radiation backgrounds ($\J$ and $F_{\rm ion}$) and the streaming velocity can either be calculated self-consistently or \textit{a priori} using a fixed UV background, as done in the example calculation in Fig. \ref{SummaryCooling}. In future work (Paper II, Nebrin et al., in preparation), we will use the model of $M_{\rm crit}$ presented in this paper as an integral part of {\sc Anaxagoras}, a physically comprehensive semi-analytical model of star formation in low-mass haloes at Cosmic Dawn. 

We note that while our model of $M_{\rm crit}$ has taken many important processes into account, we have not included X-ray feedback, which in certain regimes could potentially aid the formation of H$_2$ and hence lower $M_{\rm crit}$ \citep{Machacek2003_Xrays, Ricotti2016_Xray, Park2021}. This effect could be added in future work, but we decided not to include it here for two reasons. First, the X-ray background at very high redshifts $z \gtrsim 15$ is highly uncertain and any predictions would depend on e.g. the assumed IMF of Population III stars \citep{Ricotti2016_Xray}. Hence it is not clear whether the actual X-ray background at these high redshifts can boost the formation of H$_2$ significantly. Secondly, for $z \gtrsim 15$ the cooling threshold in most regions is strongly regulated by baryon streaming velocities \citep[not included in the analysis of][]{Ricotti2016_Xray,Park2021}, and not just the efficiency of H$_2$ formation. Thus, we expect streaming velocities to put an effective upper limit to the effectiveness of X-ray feedback on $M_{\rm crit}$ at high redshifts.

As the James Webb Space Telescope observations are opening up the extreme redshifts at which minihaloes are expected to contribute substantially to the star formation in the Universe \citep[e.g.][]{zackrisson2020bubble,qin2020tale,robertson2022discovery}, 
analytical models for their star formation properties will be needed to efficiently explore their impact. Our new model for the cooling threshold, which provides a better match to simulation results than previous models,  can be a key ingredient for such explorations.

\section*{Acknowledgements}

We thank Anna T. P. Schauer, Hajime Susa, Kenji Hasegawa, and Danielle Skinner for kindly providing simulation data shown in Figs. \ref{H2coolingfigure} and \ref{fig:J21} and for helpful discussions. ON is also indebted to Muhammad A. Latif, Shingo Hirano, Riouhei Nakatani, Stuart L. West, and Friskis Vikings for many helpful comments and stimulating discussions regarding this project. ON and GM acknowledge support from the Swedish Research Council grant 2020-04691.
Nordita is supported in part by NordForsk.

\section*{Data availability}
There are no new data associated with this article. The data points shown in various figures were extracted from previous papers and in some cases raw data from previous results were made available by the original authors.




\bibliography{bibfile}



\appendix

\section{Effect of different self-shielding fits on the cooling threshold}
\label{shieldingeffect}

\begin{figure}
    \includegraphics[width = \columnwidth]{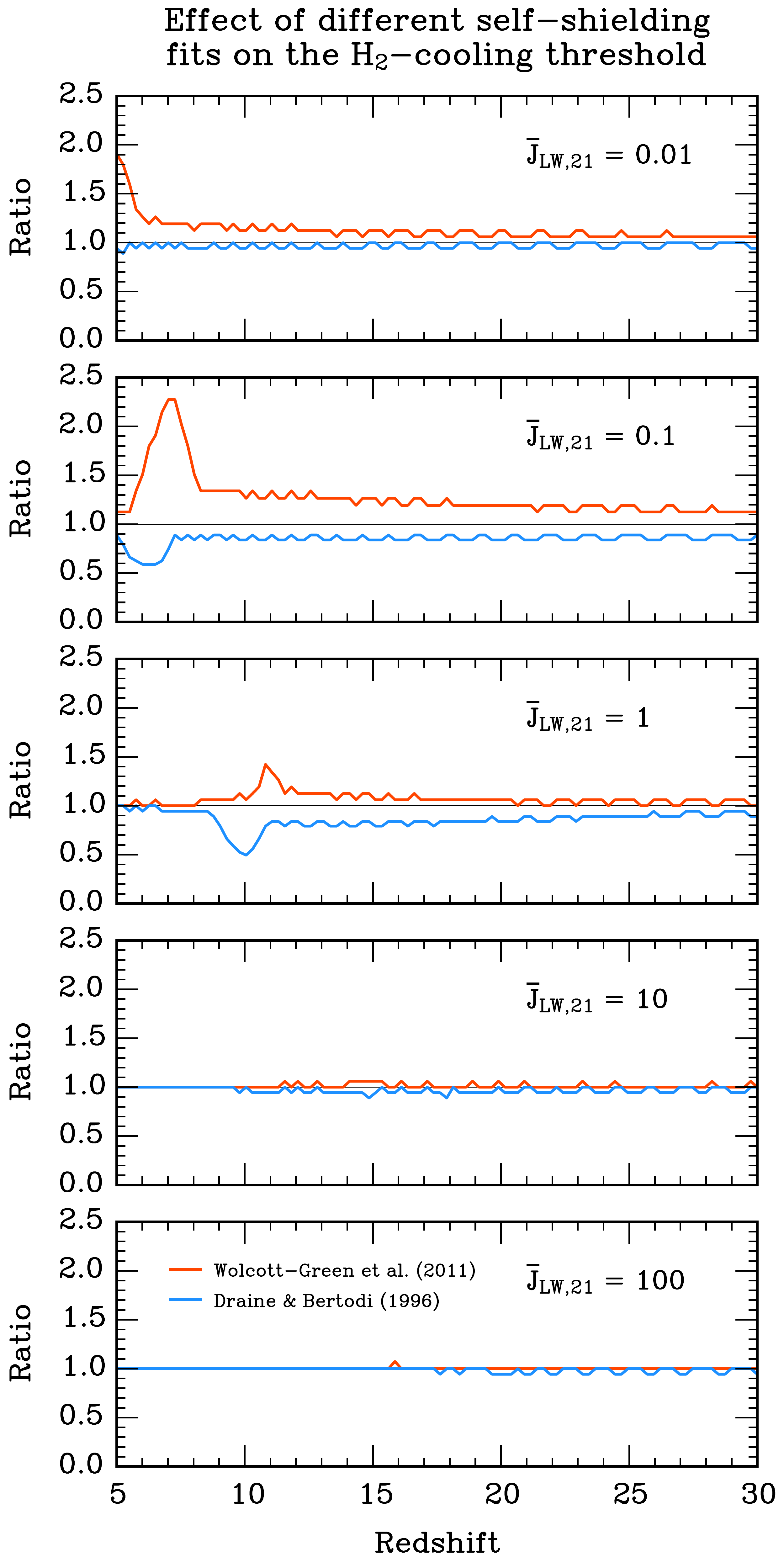}
    \caption{The ratio between the calculated halo mass cooling thresholds with either \citet{Draine1996} (blue) or \citet{WolcottGreen2011} (red), to the fiducial case in the main text that assumed the fit from \citet{WolcottGreen2019}. The panels represent different LW intensities 
    $\Bar{J}_{\rm LW,21}$.}
    \label{fig:shield}
\end{figure}

As discussed in Sec.~\ref{sec:LWfeedback}, most simulations of minihaloes that include self-shielding by H$_{2}$ utilise fits of $f_{\rm sh}$ from \citet{WolcottGreen2011} or \citet{Draine1996}. The fit by \citet{Draine1996} implicitly assumes that the gas is sufficiently cold and the rotational levels are in their ground states. It was adopted by \citet{Schauer2021} in their studies of cooling in minihaloes, and in the FIRE-3 galaxy formation simulations by \citet{FIRE3}, which aim to model H$_2$ chemistry in the early Universe (relevant e.g. for Pop III star formation in minihaloes and subsequent pre-enrichment of Ultra-Faint Dwarf galaxies).  \citet{WolcottGreen2011} updated the fit by \citet{Draine1996} to incorporate local thermal equilibrium (LTE) rotational level populations in the vibrational ground state, which they show gives rise to a weaker self-shielding. Finally, \citet{WolcottGreen2019} further updated the calculation of self-shielding by computing the full-level populations with CLOUDY \citep{Cloudy}. Their result falls somewhere between the earlier extremes of \citet{Draine1996} and \citet{WolcottGreen2011} and is the most accurate fit available.  

Since we compare our semi-analytical calculations to simulations that include different self-shielding fits, it is worth checking how these different fits affect the estimated cooling thresholds. We therefore re-calculate $f_{\rm sh}$ from both \citet{Draine1996} and \citet{WolcottGreen2011}, keeping everything else unchanged. Fig.~\ref{fig:shield} shows the results in the form of the ratio between the cooling thresholds calculated with either the \citet{Draine1996} or \citet{WolcottGreen2011} fits, to the fiducial case in the main text that uses the fit from \citet{WolcottGreen2019}. 
For redshifts $z \gtrsim 10$, the errors are typically at most $\sim \textrm{few} \times 10\%$. In particular, the fit from \citet{Draine1996} (strongest self-shielding) slightly underestimates the cooling threshold, whereas the fit from \citet{WolcottGreen2011} (weakest self-shielding) slightly overestimates the cooling threshold, as expected. The differences are small and well within the scatter in simulations, and can therefore be neglected. More significant errors up to a factor $\sim 2$ are found for low LW intensities ($\overline{J}_{\rm LW,21} \lesssim 0.1$) at redshifts $z \lesssim 10$. However, the LW background at these redshifts is expected to be higher than this, so in practice this is unlikely to lead to large errors.  

\section{The photoevaporation mass scale in the presence of a D-type ionization front}
\label{photoevaporationscales}

In this Appendix we present the mathematical details for the derivation of the mass scale $M_{\rm sh, D}$, the halo mass above which most haloes can self-shield effectively against D-type ionization fronts (see Sec. \ref{Dtypefrontmass}). We can determine $M_{\rm sh, D}$ by setting the left-hand side of Eq. (\ref{tevap_half_final}) for the photoevaporation time-scale equal to $t_{\rm ff}$, and expressing the right-hand side in terms of the virial radius $\Rvir$ of the halo of mass $M_{\rm sh, D}$:
 \begin{align}
    t_{\rm ff} ~ &= A \Rvir^{3/2} \left[ \left(\frac{B}{\Rvir^{1/2}}\right)^b + \left(\frac{C}{\Rvir^{1/4}}\right)^b \right ]^{1/b} \nonumber \\ ~ &= A \left[ B^b \Rvir^b + C^b \Rvir^{5b/4} \right ]^{1/b} \hspace{1 pt} \label{tevaproot1},
\end{align}
where
\begin{align}
    A ~ &\equiv \left( \frac{8 k_1 n_{\rm core}}{3 c_{\rm s, HII}^3} \right)^{1/2} \hspace{1 pt} , \nonumber \\
    B ~ &\equiv \frac{0.096985}{(4 k_1 F_{\rm ion}/3 c_{\rm s,HII}^2)^{1/2}} \hspace{1 pt} , \\
    C ~ &\equiv \frac{0.062978}{(4 k_1 F_{\rm ion}/3 c_{\rm s,HII}^2)^{1/4}} \nonumber \hspace{1 pt} . 
\end{align}
Rearranging Eq. (\ref{tevaproot1}) yields
\begin{equation}
    \Rvir^{5b/4} + \left( \frac{B}{C} \right)^b \Rvir^b = \left( \frac{t_{\rm ff}}{AC} \right)^b \hspace{1 pt} \label{tevaproot2}.
\end{equation}
We find that the solution of Eq. (\ref{tevaproot2}) can be approximated well by:
\begin{align}
    \Rvir ~ &\simeq \frac{t_{\rm ff}/AB}{ \left[1 + \left( \dfrac{C t_{\rm ff}^{1/4}}{A^{1/4} B^{5/4}} \right)^\gamma \right]^{4/5\gamma}} \nonumber \\
    ~ &= \frac{7.29 \hspace{1 pt} (F_{\rm ion} c_{\rm s,HII}/n_{\rm core})^{1/2} t_{\rm ff}}{ \left[1 + 1.15^\gamma \left( \dfrac{k_1^2 F_{\rm ion}^3 t_{\rm ff}^2}{n_{\rm core} c_{\rm s,HII}^3} \right)^{\gamma/8}  \right]^{4/5\gamma}} \hspace{1 pt} \label{Rvir_root},
\end{align}
with $\gamma = 2.6$. This approximation has the correct asymptotic limits and a maximum relative error of $0.52\%$.\footnote{The approximate solution in Eq. (\ref{Rvir_root}) was found as follows. First, we note that Eq. (\ref{tevaproot2}) has the solution $\Rvir \simeq t_{\rm ff}/AB$ when $B/C$ becomes sufficiently large. Thus, it is convenient to define $x \equiv \Rvir/(t_{\rm ff}/AB)$. This yields $p^b x^{5b/4} + x^b - 1 = 0$ with $p \equiv C t_{\rm ff}^{1/4}/A^{1/4} B^{5/4}$. We have thus reduced the problem to one with a single free parameter $p$. The solutions for small and large $p$ are $x \simeq 1$ and $x \simeq p^{-4/5}$, respectively. Comparison with a numerical solution for $x$ show that the approximation $x \simeq 1/(1 + p^\gamma)^{4/5\gamma}$ with $\gamma = 2.6$ is accurate to within $0.52\%$ for all $p$.} The corresponding halo mass $M_{\rm sh, D}$ is given in Eq. (\ref{M_shD}).



\label{lastpage}
\end{document}